\documentclass[%
 reprint,superscriptaddress,
 amsmath,amssymb,
 aps,
]{revtex4-1}
\usepackage[utf8]{inputenc}
\usepackage{ragged2e}
\usepackage{graphicx}
\usepackage{dcolumn}
\usepackage{bm}
\usepackage{hyperref}
\usepackage{colortbl}
\usepackage{verbatim}
\usepackage{nicefrac}
\usepackage{hhline}
\usepackage{}
\usepackage[footnote]{acronym}
\newcommand{\rn}[2]{$^{#1}$#2}
\newcommand{\mgi}{$^{23}$Mg$^+$\,}
\newcommand{\cai}{$^{39}$Ca$^+$\,}
\newcommand{\nai}{$^{23}$Na$^+$\,}
\newcommand{\mg}{$^{23}$Mg}

\newcommand{\D}{$D$\,}
\newcommand{\bt}{$\beta$\,}
\newcommand{\jyu}{the University of Jyv\"askyl\"a}
\newcommand{\E}[1]{$10^{#1}$}
\usepackage{subcaption}
\usepackage{amsmath}
\usepackage[dvipsnames]{xcolor}
\usepackage[footnote]{acronym}
\usepackage[backgroundcolor=white,bordercolor=orange]{todonotes}

\begin{document}
\preprint{AIP/123-QED}
\title{Performance of the MORA Apparatus for Testing Time-Reversal Invariance in Nuclear Beta Decay}
\author{N. Goyal}
\altaffiliation[Also at ]{%
Synchrotron SOLEIL, 91190 Saint-Aubin, France
}
\affiliation{%
GANIL, CEA/DRF  - CNRS/IN2P3, Bd H. Becquerel, 14076 Caen, France
}%
\author{A.~Singh}
\altaffiliation[Also at ]{%
IRSN/LMRE, Rue du Belvédère – 91400 Orsay, France 
}
\affiliation{%
GANIL, CEA/DRF  - CNRS/IN2P3, Bd H. Becquerel, 14076 Caen, France
}%

\author{S. Daumas-Tschopp}
\affiliation{%
Université de Caen Normandie, ENSICAEN, CNRS/IN2P3, LPC Caen UMR6534, 14050 Caen, France
}%
\author{L. M. Motilla Martinez}
\affiliation{%
GANIL, CEA/DRF  - CNRS/IN2P3, Bd H. Becquerel, 14076 Caen, France
}%
\affiliation{%
University of Jyvaskyla, Department of Physics, Accelerator Laboratory, P.O. Box 35(YFL) FI-40014 University of Jyvaskyla, Finland
}
\author{G. Ban}
\affiliation{%
Université de Caen Normandie, ENSICAEN, CNRS/IN2P3, LPC Caen UMR6534, 14050 Caen, France
}%
\author{V. Bosquet}
\affiliation{%
Université de Caen Normandie, ENSICAEN, CNRS/IN2P3, LPC Caen UMR6534, 14050 Caen, France
}%
\author{J. F. Cam}
\affiliation{%
Université de Caen Normandie, ENSICAEN, CNRS/IN2P3, LPC Caen UMR6534, 14050 Caen, France
}%
\author{P. Chauveau}
\affiliation{%
GANIL, CEA/DRF  - CNRS/IN2P3, Bd H. Becquerel, 14076 Caen, France
}%
\author{S. Chinthakayala}
\affiliation{%
GANIL, CEA/DRF  - CNRS/IN2P3, Bd H. Becquerel, 14076 Caen, France
}%
\affiliation{%
University of Jyvaskyla, Department of Physics, Accelerator Laboratory, P.O. Box 35(YFL) FI-40014 University of Jyvaskyla, Finland
}
\author{G. Frémont}
\affiliation{%
GANIL, CEA/DRF  - CNRS/IN2P3, Bd H. Becquerel, 14076 Caen, France
}%
\author{R.P. De Groote}
\affiliation{%
Instituut voor Kern- en Stralingsfysica, KU Leuven, B-3001 Leuven, Belgium
}
\author{F. de Oliveira Santos}
\affiliation{%
GANIL, CEA/DRF  - CNRS/IN2P3, Bd H. Becquerel, 14076 Caen, France
}%
\author{T. Eronen}
\affiliation{%
University of Jyvaskyla, Department of Physics, Accelerator Laboratory, P.O. Box 35(YFL) FI-40014 University of Jyvaskyla, Finland
}
\author{A. Falkowski}
\affiliation{%
IJCLab, 15 Rue Georges Clemenceau, 91400 Orsay, France
}
\author{X. Fléchard}
\affiliation{%
Université de Caen Normandie, ENSICAEN, CNRS/IN2P3, LPC Caen UMR6534, 14050 Caen, France
}%
\author{Z. Ge}
\affiliation{%
University of Jyvaskyla, Department of Physics, Accelerator Laboratory, P.O. Box 35(YFL) FI-40014 University of Jyvaskyla, Finland
}
\author{M. González–Alonso}
\affiliation{%
IFIC, Universitat de València - CSIC, 46980 Paterna, Spain
}
\author{H. Guérin}
\affiliation{%
GANIL, CEA/DRF  - CNRS/IN2P3, Bd H. Becquerel, 14076 Caen, France
}%
\author{L. Hayen}
\affiliation{%
Université de Caen Normandie, ENSICAEN, CNRS/IN2P3, LPC Caen UMR6534, 14050 Caen, France
}%
\author{A. Jaries}
\affiliation{%
University of Jyvaskyla, Department of Physics, Accelerator Laboratory, P.O. Box 35(YFL) FI-40014 University of Jyvaskyla, Finland
}
\author{M. Jbayli}
\affiliation{%
GANIL, CEA/DRF  - CNRS/IN2P3, Bd H. Becquerel, 14076 Caen, France
}
\author{A. Jokinen}
\affiliation{%
University of Jyvaskyla, Department of Physics, Accelerator Laboratory, P.O. Box 35(YFL) FI-40014 University of Jyvaskyla, Finland
}
\author{A. Kankainen}
\affiliation{%
University of Jyvaskyla, Department of Physics, Accelerator Laboratory, P.O. Box 35(YFL) FI-40014 University of Jyvaskyla, Finland
}
\author{B. Kootte}
\affiliation{%
University of Jyvaskyla, Department of Physics, Accelerator Laboratory, P.O. Box 35(YFL) FI-40014 University of Jyvaskyla, Finland
}
\author{R. Kronholm}
\affiliation{%
University of Jyvaskyla, Department of Physics, Accelerator Laboratory, P.O. Box 35(YFL) FI-40014 University of Jyvaskyla, Finland
}
\author{N. Lecesne}
\affiliation{%
GANIL, CEA/DRF  - CNRS/IN2P3, Bd H. Becquerel, 14076 Caen, France
}%
\author{Y. Merrer}
\affiliation{%
Université de Caen Normandie, ENSICAEN, CNRS/IN2P3, LPC Caen UMR6534, 14050 Caen, France
}%
\author{V. Morel}
\affiliation{%
GANIL, CEA/DRF  - CNRS/IN2P3, Bd H. Becquerel, 14076 Caen, France
}%
\author{M. Mougeot}
\affiliation{%
University of Jyvaskyla, Department of Physics, Accelerator Laboratory, P.O. Box 35(YFL) FI-40014 University of Jyvaskyla, Finland
}
\author{G. Neyens}
\affiliation{%
Instituut voor Kern- en Stralingsfysica, KU Leuven, B-3001 Leuven, Belgium
}
\author{J. Perronnel}
\affiliation{%
Université de Caen Normandie, ENSICAEN, CNRS/IN2P3, LPC Caen UMR6534, 14050 Caen, France
}%
\author{M. Reponen}
\affiliation{%
University of Jyvaskyla, Department of Physics, Accelerator Laboratory, P.O. Box 35(YFL) FI-40014 University of Jyvaskyla, Finland
}
\author{A. Raggio}
\affiliation{%
University of Jyvaskyla, Department of Physics, Accelerator Laboratory, P.O. Box 35(YFL) FI-40014 University of Jyvaskyla, Finland
}
\author{S. Rinta-Antila}
\affiliation{%
University of Jyvaskyla, Department of Physics, Accelerator Laboratory, P.O. Box 35(YFL) FI-40014 University of Jyvaskyla, Finland
}
\author{A. Rodriguez – Sanchez}
\affiliation{%
IFIC, Universitat de València - CSIC, 46980 Paterna, Spain
}
\author{N. Severijns}
\affiliation{%
Instituut voor Kern- en Stralingsfysica, KU Leuven, B-3001 Leuven, Belgium
}
\author{J. C. Thomas}
\affiliation{%
GANIL, CEA/DRF  - CNRS/IN2P3, Bd H. Becquerel, 14076 Caen, France
}%
\author{C. Vandamme}
\affiliation{%
Université de Caen Normandie, ENSICAEN, CNRS/IN2P3, LPC Caen UMR6534, 14050 Caen, France
}%
\author{S. Vanlangendonk}
\affiliation{%
Instituut voor Kern- en Stralingsfysica, KU Leuven, B-3001 Leuven, Belgium
}
\author{V. Virtanen}
\affiliation{%
University of Jyvaskyla, Department of Physics, Accelerator Laboratory, P.O. Box 35(YFL) FI-40014 University of Jyvaskyla, Finland
}
\author{E. Liénard}
\affiliation{%
Université de Caen Normandie, ENSICAEN, CNRS/IN2P3, LPC Caen UMR6534, 14050 Caen, France
}%
\author{I. D. Moore}
\affiliation{%
University of Jyvaskyla, Department of Physics, Accelerator Laboratory, P.O. Box 35(YFL) FI-40014 University of Jyvaskyla, Finland
}
\author{P. Delahaye}
\affiliation{%
GANIL, CEA/DRF  - CNRS/IN2P3, Bd H. Becquerel, 14076 Caen, France
}%
\date{\today}
%
\begin{abstract}
The MORA experimental setup is designed to measure the triple-correlation \D parameter in nuclear beta decay. The \D coefficient is sensitive to possible violations of time-reversal invariance. The experimental configuration consists of a transparent Paul trap surrounded by a detection setup with alternating \bt and recoil-ion detectors. The octagonal symmetry of the detection setup optimizes the sensitivity of positron-recoil-ion coincidence rates to the \D correlation, while reducing systematic effects. MORA utilizes an innovative in-trap laser polarization technique. The design and performance of the ion trap, associated beamline elements, lasers and \bt and recoil-ion detectors, are presented. Recent progress towards the polarization proof-of-principle is described. 
\end{abstract}
\maketitle
\section{Introduction}
\label{sec:Introduction}
The Standard Model (SM) of particle physics currently provides the most compelling description of the fundamental interactions between elementary particles. Despite its success, several profound questions remain unanswered. Among those, the SM requires at least 19 arbitrary parameters, the nature of dark energy and dark matter remains unknown, and the origin for the imbalance between matter and antimatter in the universe is still unexplained. A promising avenue for probing New Physics (NP) beyond the Standard Model is through precision measurements in nuclear \bt decay. \par 
 
The study of \bt-decay processes provides a unique laboratory for scrutinizing the weak interaction, one of the four fundamental forces, with unparalleled precision.
Historically, the investigation of \bt decay has led to critical developments in our understanding of weak interactions, from the discovery of parity violation to the formulation of the {\it V-A} theory. Recent advances in experimental techniques and detector technologies have significantly enhanced the precision with which \bt decay parameters can be measured. These advancements open new windows to search for subtle deviations from the predictions of the SM, which may indicate the presence of NP.  As such, \bt decay experiments offer a complementary approach to experiments which directly probe energy scales of NP through particle collisions at high energies. The complementarity and interplay of these approaches are now very well established thanks to modern Effective Field Theories. With sensitivities reaching the $\sim$\E{-3} level on traces of non-standard Wilson coefficients in the effective Hamiltonian of this low-energy process, constraints on interactions involving new particles beyond the TeV regime, such as Leptoquarks or right-handed W bosons, can compete with limits obtained at the Large Hadron Collider (LHC) \cite{GonzalezAlonso2013, GonzalezAlonso2019, Falkowski2021}. \par 
The Matters Origin from RadioActivity (MORA) project searches for NP by studying the \D correlation in nuclear \bt decay, which occurs in mixed Fermi and Gamow-Teller transitions of spin~polarized nuclei~\cite{Delahaye2019}. The \D correlation is a triple correlation between the spin of the decaying nucleus, the momentum of the emitted electron or positron, and the momentum of the emitted neutrino. It is particularly interesting because it is sensitive to time-reversal symmetry violation, not only in yet unobserved scalar and tensor currents, but also in well-established V, A currents.  Any observed T-violating effects are a smoking gun for NP and provide insight into new CP-violation mechanisms, which are necessary to explain the baryon asymmetry in the universe \cite{Sakharov1991}. \par
The particular case of the sensitivity of \D to NP was recently reviewed by coauthors of this article \cite{Falkowski2022}. While the search for NP in the framework of specific albeit popular NP Models is constrained to the \E{-5} level by Electric Dipole Moments (EDMs), pion decay and the LHC, a manifestation of CP violation at the \E{-4} level is not excluded by a direct measurement of $D$. The most constraining measurement to date has been obtained from neutron decay by the emiT experiment:$D_n<2\cdot 10^{-4}$ \cite{Chupp2012, Mumm2011}. The MORA experimental program is therefore organized in two phases: the first phase, conducted at the Ion Guide Isotope Separation On-Line (IGISOL) facility \cite{Aysto:2014xwa} at \jyu, Finland, aims to obtain a sensitivity to a non-zero \D correlation competitive with the emiT experiment; the second phase, carried out at GANIL in the future DESIR facility, targets a sensitivity of the order of $\sim$\E{-5} \cite{Delahaye2019}, sufficient to start probing NP from specific models. \par
To acquire the required statistics for the ambitious goals of the experiment, MORA makes use of an innovative technique, combining the high efficiencies of ion trapping and laser polarization.  We present here the current status of the MORA experimental setup at IGISOL.  The following sections describe successively the experimental setup, the commissioning of the beamline, the ion trap and its performances, and the characterization of the detection setup. The calculation of the sensitivity factor on $D$, that depends on the phase space selected by the detection setup of MORA (parameter $\delta$ introduced in Eq. 5 of ref. \cite{Delahaye2019}), and the evaluation of possible systematic effects are then presented. The last section concludes on the results of the latest experimental progress to demonstrate the laser polarization technique.
\section{The MORA apparatus}
The MORA experimental setup is a novel apparatus that has been designed and built to store and polarize radioactive ions and observe their subsequent decay. With this apparatus, \mgi and \cai  ions are the most promising candidates for the \D correlation measurement \cite{Delahaye2019}. The whole MORA setup consists of an injection beam line, a transparent Paul trap and a detection setup.  The injection beam line decelerates the ions delivered by the IGISOL RFQ cooler buncher to allow their efficient capture in the transparent Paul trap \cite{Benali2020, Delahaye2019a}. The latter permits the in-situ polarization of the ion cloud. The detection setup surrounding the trap is optimized to maximize solid-angle coverage for positrons and recoil ions emitted after \bt decay.
\subsection{MORA at IGISOL}
\label{sec:production}
\begin{figure*}
    \centering
     \includegraphics[width=0.7\textwidth]{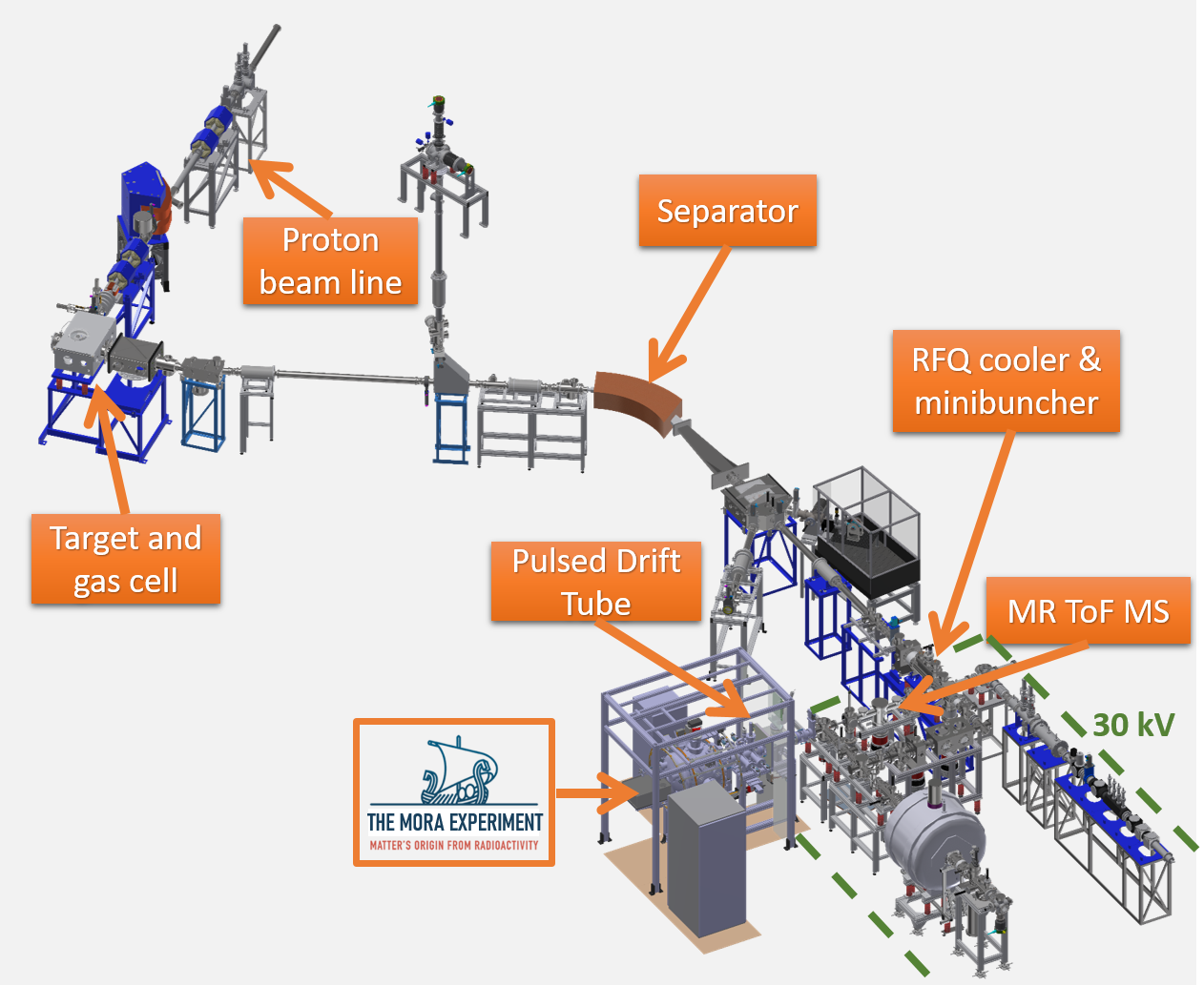}
    \caption{\small Sketch of the production system and IGISOL hall featuring the different components important to MORA. The dashed line shows the footprint of the high voltage cage.}
    \label{fig:IGISOL_Hall}
\end{figure*}

Figure~\ref{fig:IGISOL_Hall} shows the integration of the MORA apparatus within the IGISOL facility \cite{Aysto:2014xwa}. At IGISOL, low-energy radioactive ion beams are produced via light-particle (p, d, $^3$He, $^4$He) and heavy-ion induced reactions in a gas cell. These reactions include fission, used in the majority of experiments, as well as transfer or fusion-evaporation reactions, of particular interest for MORA.  Because of charge recombination with the gas and impurities in the cell, most of the reaction products stopped in the gas are eventually delivered from the cell as singly charged ions. A sextupole ion guide \cite{KARVONEN20084794} permits their collection as a beam before acceleration to 30~keV, and subsequent mass separation thanks to a dipole magnet. In the IGISOL hall, the radioactive beams can be cooled by the IGISOL RFQ cooler in a High Voltage (HV) cage, which has been recently upgraded with a minibuncher to enable bunching of ions with a time spread $< 100$~ns \cite{VIRTANEN2025170186}. The minibuncher has been developed for the Multi-Reflection Time-of-Flight Mass Spectrometer (MR-ToF-MS) of IGISOL, allowing for a fast separation of isobars and mass measurement of short-lived isotopes. Ions are transported to different setups at an energy of $\sim$ 2 keV (e.g. to the JYFLTRAP Penning trap \cite{Eronen2012} or RAPTOR \cite{Kujanpaa2023}). \par
For practical reasons, including protecting fragile electronics and detection setup from sparks, and avoiding possible sources of noise related to the delicate definition of reference potentials, the MORA apparatus is standing just beyond the HV cage downstream from the MR-ToF-MS. MORA therefore requires a particular beam manipulation for the decelerating and trapping of ions to sub-eV energies, as is detailed in the next section. So far the beam production and manipulation efforts for MORA have been limited to \mgi ions, as those only require one laser frequency for their polarization, compared to two for \cai ions. The laser light required for the polarization of \mgi ions is prepared on a laser table neighbouring the MORA setup, also used for RAPTOR campaigns. Circular polarization of the laser beam and control over its handedness required for MORA is done on a small platform (80~cm x 50~cm) attached to the trap chamber of MORA (see section \ref{sec:Laser_polarisation}). The production of \mgi ions in the IGISOL gas cell is described in section \ref{sec:Experimental_results}. 

\subsection{MORA beam line}
\label{sec:Line}

\begin{figure*}
    \centering
     \includegraphics[width=1.0\textwidth]{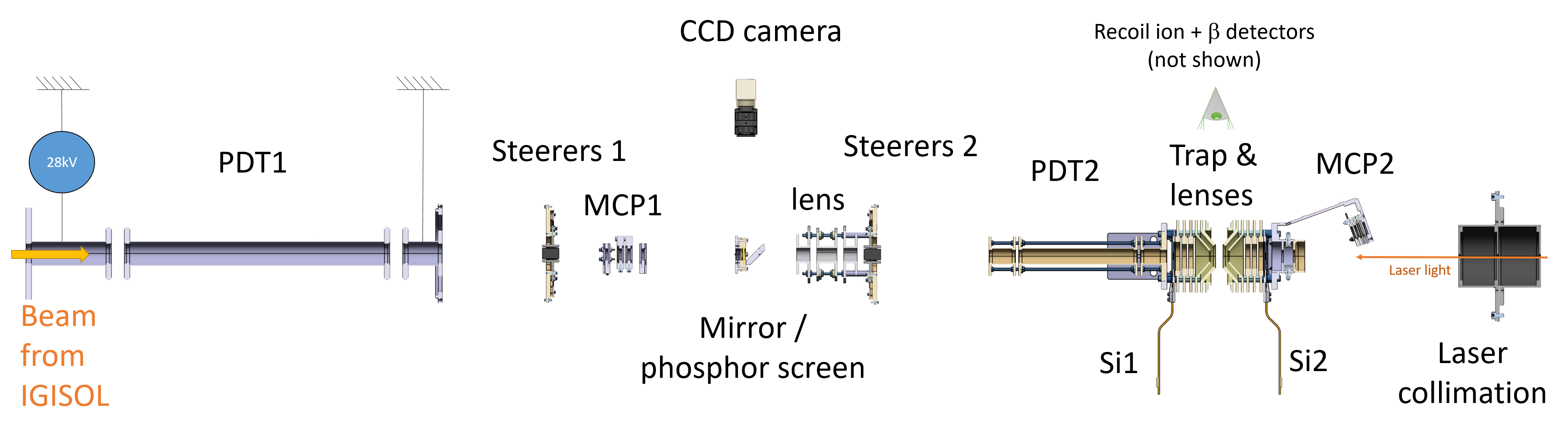}
    \caption{\small Simplified view of the injection line, trap and ion counting systems.}
    \label{fig:Line}
\end{figure*}

To enable efficient deceleration and capture of ions in the trap of MORA, the injection beam line has been originally designed using SIMION \cite{SIMION} simulations . The mechanical components of MORA, i.e. ion optics, electrodes of the ion trap, trap and injection chambers, and the detection frames have been conceived and manufactured in-house, by the LPC Caen workshop, to enable an alignment precision of $\leq$ 100~µm. Special care was particularly taken for the support of detectors surrounding the trap described in section \ref{sec:Detection}. The injection beam line is shown in Fig.~\ref{fig:Line}, together with the ion trap. It contains two Pulsed Drift Tubes (PDTs) for the deceleration of ions, ion optics for steering and focusing, and different diagnostics for ion counting and laser alignment. 
 
 \subsubsection{Deceleration of ions} 
The PDTs enable the progressive deceleration of the bunches delivered by the minibuncher for their efficient capture in the trap. They consist of a central cylindrical electrode framed by two shorter ones at the reference potential of the beam line. The central electrode of PDT1 is half contained in the HV cage of IGISOL and half in the grounded part of the injection beam line (the HV and injection beam lines are connected by a ceramic break surrounding the PDT central electrode). The upstream reference potential is therefore about +28 kV to ground (reference potential of the HV cage), while the downstream one is ground. For PDT2 the reference potential is simply ground on both sides. The length of the central electrode for both PDTs has been chosen in order to accomodate with a safety margin (factor $\sim 3$) the spatial distribution of \mgi ion bunches of 1 µs: 40 cm for PDT1, for accepting bunches of 2 keV, and 18 cm for PDT2, for accepting 150 eV bunches. The principle of the deceleration of ions in the PDTs is illustrated in Fig. \ref{Fig:PDT}. When the ion bunch is crossing the center of the drift tube, the voltage of the tube, playing here the role of a Faraday cage, is switched from a positive value to a negative one. The ion's motion is therefore not affected by the sudden change of potential until it reaches the end of the drift tube, where it is eventually decelerated. The balancing of the positive and negative voltages enables control on beam focussing, similar to that obtained by tuning the RF phase of accelerating cavities. \par
\begin{figure}[h!]
    \begin{center}
        \includegraphics[scale=0.45,angle = 0,keepaspectratio]{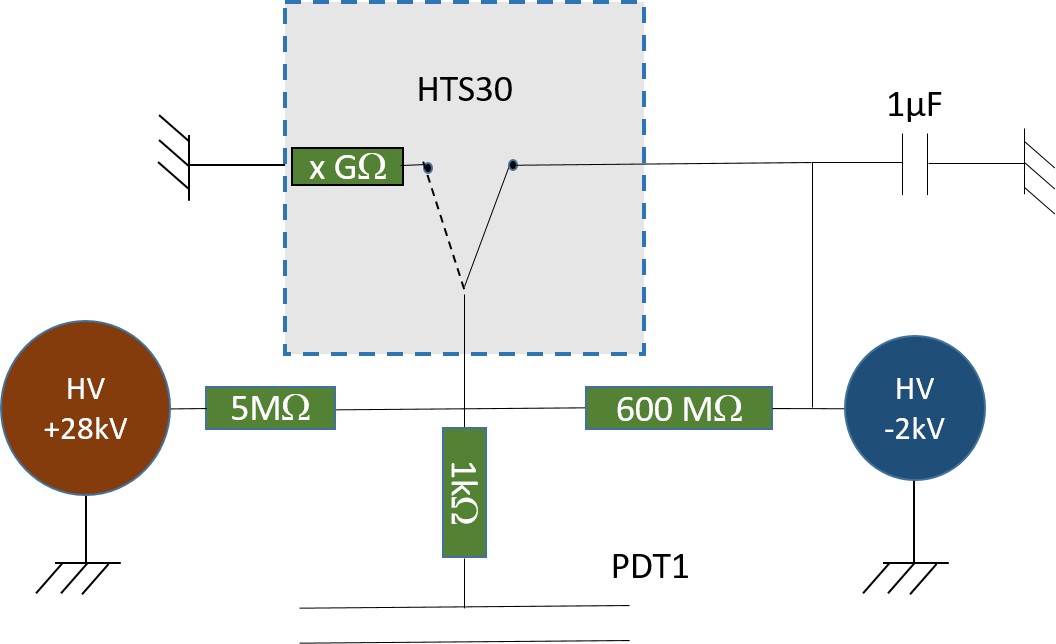}
    \end{center}
    \captionsetup{singlelinecheck=off}
\caption{\small Sketch of the switching circuit. The HTS30 switch from Behlke has an impedance of several G$\Omega$ to ground.}
\label{fig:HT_switch}
\end{figure}
\begin{figure*}[!htp]
    \begin{center}
        \includegraphics[scale=0.55,angle = 0,keepaspectratio]{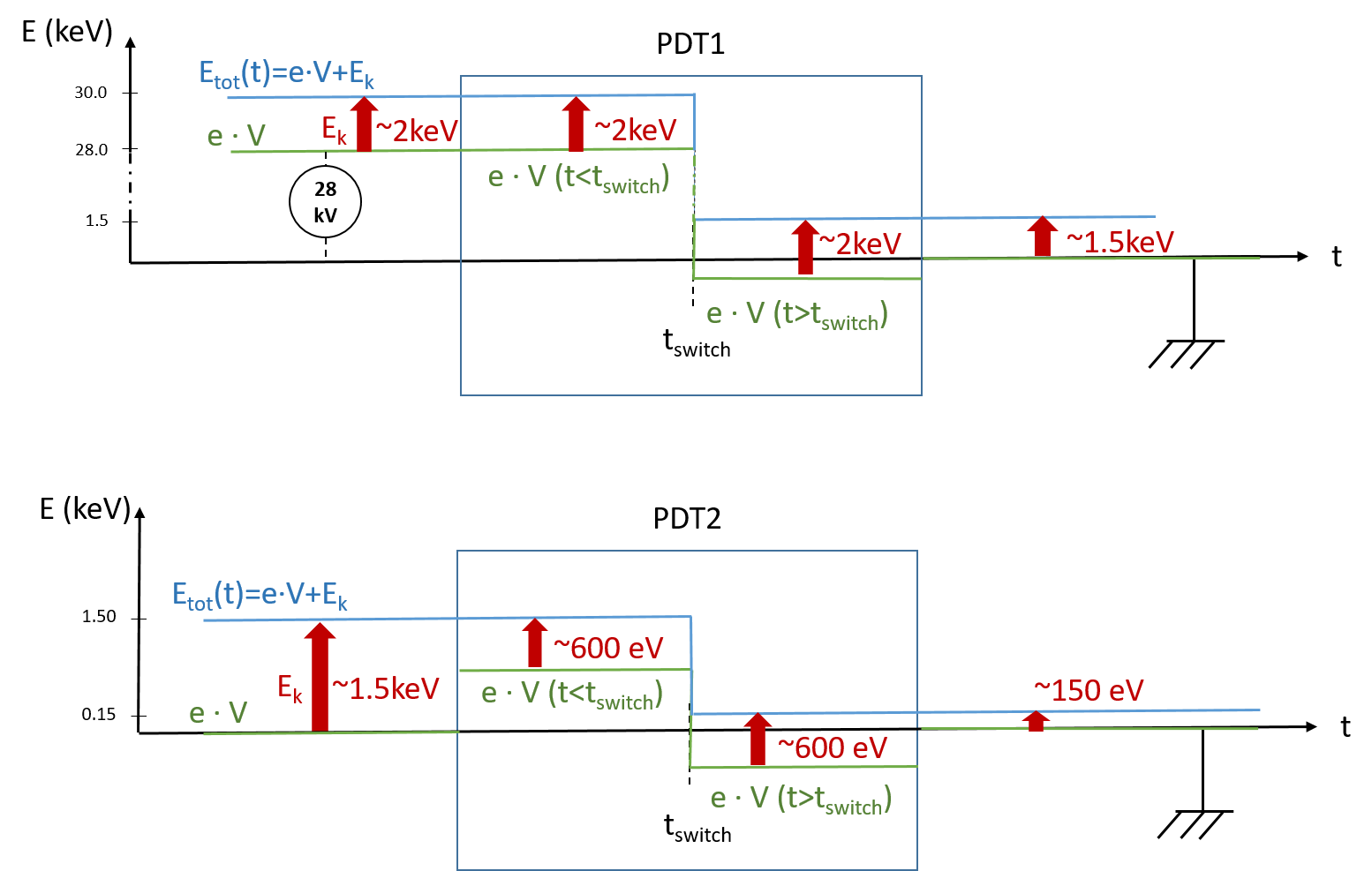}
    \end{center}
 \captionsetup{singlelinecheck=off}
\caption{\small Principle of deceleration in PDT1 and PDT2. The total, kinetic and potential energies (resp. E$_{tot}$, E$_k$ and e$\cdot$V) of ions passing through the PDTs are plotted as a function of their time-of-flight t. }
\label{Fig:PDT}
\end{figure*}

PDT1 reduces the kinetic energy of ions crossing the HV cage towards the injection beam line by $\sim$ 500~eV . Its potential is switched from approximately 28~kV (the reference potential of the HV cage) to -500~V relative to ground, using a 30~kV Behlke switch from the HTS series, together with Spellman +30~kV / -2~kV power supplies. The quality of this switching is critical for MORA, as it must avoid generating energy dispersion or shifts larger than $\sim$ 10 eV, which would reduce the efficiency of the ion capture in the trap. While achieving a sufficiently short switching time - less than 1 µs compared to the ions' transit time  of $\simeq$ 3 µs - has never been an issue, maintaining voltage stability before and after switching, when ions enter or exit the PDT, has proven to be a challenging task. Multiple sources of large, rapid fluctuations (µs/ms), as well as longer-term drifts and variations (min/h), had to be identified and addressed. The key measures taken to eliminate these sources are listed below:
\begin{itemize}
\item The initial configuration of our switch included two HTS30 modules from Behlke in cascade to allow for a possible usage of the switch with up to 60~kV. The switching circuit was simplified, in order to avoid problems of desynchronisation of the HTS modules, as indicated by Behlke. The final version of the switching circuit is shown in Fig.~\ref{fig:HT_switch}.
\item The impedances of this circuit had to be carefully selected so as to produce stable pulsing. In particular the rather small 5 M$\Omega$ resistor was chosen to reduce the dependence on the positive voltage to possible impedance fluctuations of the HTS switch. The 1 k$\Omega$ resistor was found suitable for optimizing trapping efficiencies. With such a resistor a fast switching of $\sim$ 200 ns could be obtained, while reducing voltage oscillations after switching. 
\item The resistors and connections to the positive and switch output were covered with kapton to avoid  micro-sparks. These microsparks, observed experimentally, were inducing erratic energy variations, up to 100 eV. 
\item The logical signal triggering the switch and the +24V switch supply had to be shielded against the EM noise generated by the switching to avoid unwanted retriggering or switch malfunction. 
\item The electrical connection of the positive HV supply to the 5 M$\Omega$  had to be redone several times to ensure firm contact, and prevent microsparks during the switch operation.
\item Grounding loops between the HV power supplies and the switch circuit generated unwanted noise and had to be eliminated.
\end{itemize}
The recent elimination of all the sources of instability resulted in a drastic improvement in the trapping efficiencies, presented in section \ref{sec:Trap}). \par
PDT2 requires the switching of more modest voltages. It is therefore less critical and has always been well-functioning. PDT2 decelerates ions further down to 150 eV, just before the ion trap. PDT2 voltages are produced by a 2kV switch from CGC, fed by two ISEG +1.5kV and -0.5kV power supplies. The switching is very rapid ($\sim$ 10 ns) and produces negligible fluctuations after 1 µs, when ions exit from the PDT2. \par 
The injection beam line also includes beam optics to align and focus the beam through the PDTs; two sets of x and y steerers formed of four electrodes are placed just after PDT1 and just before PDT2. An Einzel lens is placed in the middle of the steerers to focus the beam properly before entering PDT2. 

\subsubsection{Diagnostics for the deceleration of ions}
\begin{figure}
    \begin{center}
        \includegraphics[scale=0.27,angle = 0,keepaspectratio]{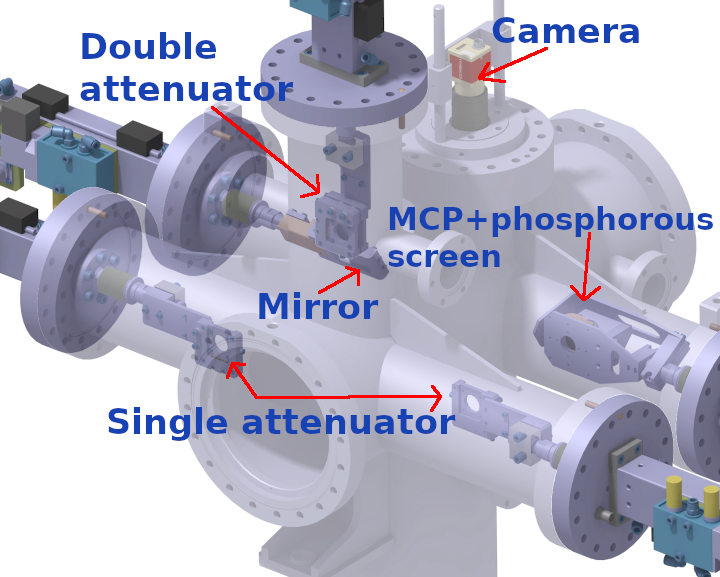}
    \end{center}
    \captionsetup{singlelinecheck=off}
\caption{\small Diagnostics of the injection beamline displayed on the user interface.}
\label{fig:U_interface}
\end{figure}

Three sets of attenuators, each consisting of electroformed meshes with 10$\%$ transmission and mounted on pneumatic actuators, help regulate the ion flux. This enables ion-by-ion detection using a removable Micro-Channel Plate detector (MCP1) and associated phosphor screen, also mounted on an actuator (Fig.~\ref{fig:U_interface}).
The phosphor screen can be used to get a qualitative display of the beam shape and position. It is read by a \textsc{ccd} camera wrapped with a piece of black fabric to get rid of external light. Alternatively, the MCP1 and phosphor screen can be replaced by a mirror to monitor the alignment of the laser beam used to polarize the ion cloud (see section \ref{sec:Laser_polarisation}). 

MCP1 also features a set of two meshes: one at ground potential with a 90$\%$ transmission efficiency and another one at a variable potential, with an additional 10$\%$ transmission factor. This retardation potential can be used to measure the energy of the beam with an energy resolution of a few eV, to be compared to an expected dispersion of $\sim$ 10 eV after the minibuncher, more or less matching the acceptance of the trap of MORA. With all the sets of attenuation meshes, up to a few $10^5$ ions per bunch can safely be detected by such an ion counting system. The MCP1 calibration is done using fast cycles and comparing the ion counting rate with measurement of currents on FCs situated in the HV cage. Time-of-flight of ions are recorded and analyzed thanks to the FASTER acquisition \cite{FASTERwebsite}, developed by LPC Caen, and also used for the detection system described in section \ref{sec:Detection}.
\subsection{Ion trap}
\label{sec:Trap}

The trap of MORA is situated at the end of the injection beam line and in the center of the detection setup. Its open geometry has been adapted from LPCTrap \cite{Delahaye2019a}, and further optimized by simulations to enlarge the volume in which the potential is harmonic \cite{Benali2020}. As a result, an increase of 27\% of the radial acceptance and an increase of 50\% of the ion capacity is expected, for the same RF voltage and for a comparable characteristic dimension, usually referred to as $r_0$ \cite{March_Todd}. 

\subsubsection{Trap characteristics}
\begin{figure}[h!]
    \begin{center}
        \includegraphics[scale=0.55,angle = 0,keepaspectratio]{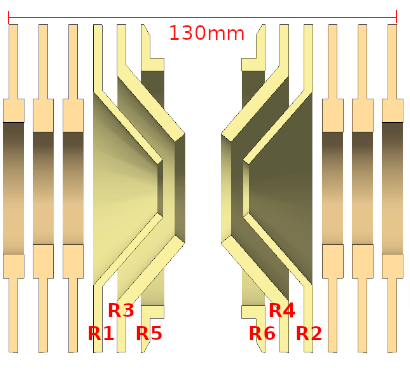}
    \end{center}
    \captionsetup{singlelinecheck=off}
\caption{\small Schematic view of the MORA trap displayed on the slow control user interface. 
}
\label{fig:Trap_scheme}
\end{figure}

The trap is composed of six independent ring-shaped electrodes labeled R1 to R6 in Fig.~\ref{fig:Trap_scheme}. R3 and R4 are pulsed electrodes used for respectively ejecting and injecting ions from / into the trap. During trapping, these electrodes are grounded, and the RF voltage is applied on R1 and R2. Using standard notations \cite{March_Todd}, R1 and R2 act as end caps, while R3 and R4 serves as the ring electrode. R5 and R6 are correcting electrodes, always at ground. This configuration minimizes RF disturbances affecting recoil ions produced by 
\bt decay as they travel to the detectors. The trap is enclosed by two sets of einzel lenses, which optimize the focus of the ion bunch for injection and extraction. These einzel lenses are switched off during trapping to minimize their disturbance on the RF field.

The RF is generated by a generator from Keysight and amplified by an amplifier from Falco Systems, capable of delivering a nearly perfect sine wave of $\sim$200V peak-to-peak on a wide range of frequencies, from a few 100kHz to a few MHz. \newline

A buffer gas of \rn{4}{He} is injected at a pressure of approximately \E{-5}~mbar during trapping for ion cooling and reducing the Doppler effect for an efficient laser polarization. 

\subsubsection{Diagnostics for ion trapping}
On the side  opposite the injection line, a small electrostatic deflector enables the off-axis detection and counting of the ions of the cloud on MCP2. MCP2 is very similar to MCP1. It is also equipped with a retardation mesh to measure the energy dispersion of decelerated ions passing through the ion trap. This diagnostic is used for the optimization of the PDT2 and R4 stopping voltages applied to capture the ions into the trap. Ion counting enables the determination of the performances of the trap, in terms of efficiency and ion trapping lifetime. It is also used on-line for monitoring the number of trapped ions after each cycle of injection - trapping - extraction. So far, cycles of 11~s, corresponding to $\sim$ 1 half-life of \mg\, and consisting of 6~s of trapping and 5~s of background measurement were used.  During the on-line operation, all attenuation mesh  in the injection beam line are removed (Fig. \ref{fig:U_interface}).  The attenuation factor is therefore limited to that of the mesh used for the energy measurement, directly attached to MCP2 (factor of $\sim$10). With typical trapped ion numbers of a few \E{4}, the extraction of the ion cloud results in a few E{3} ions reaching MCP2 within 100~ns. The number of trapped ions is therefore not measured using event-by-event detection, but roughly estimated by integrating cycle per cycle the total charge deposited and amplified by MCP2. With such technique, the precision on the ion number is rather modest, of the order of $50\%$.   
\subsubsection{Performances: Trapping efficiency and trapping lifetime}
\begin{figure}[h!]
    \begin{center}
        \includegraphics[scale=0.45,angle = 0,keepaspectratio]{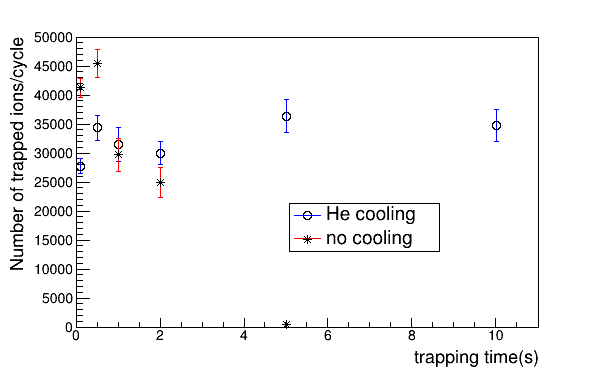}
    \end{center}
    \captionsetup{singlelinecheck=off}
\caption{\small Number of ions as a function of trapping time, measured with MCP2 with and without He cooling. The statistical errors are inflated by a factor two to take into account variations in the transport efficiencies from the IGISOL minibuncher to the trap and from the trap to MCP2. }
\label{fig:trapping_hl}
\end{figure}

\begin{figure}[h!]
    \begin{center}
      \includegraphics[scale=0.36,angle =0, keepaspectratio]{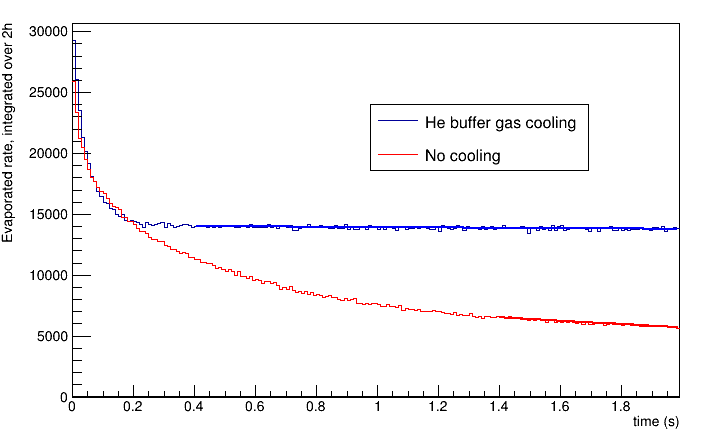}
    \end{center}
    \captionsetup{singlelinecheck=off}
\caption{\small Rates of evaporation of the \nai cloud as measured with the recoil ion detectors. The lines superimposed on the data are exponential chi-squared fits with p-values of 51$\%$ with cooling and 30$\%$ without. See text for more details. }
\label{fig:evaporation}
\end{figure}

Preliminary results obtained with a stable alkali ion source delivering a 2 keV \nai ion beam at LPC Caen and reported in \cite{dauma2023} have generally improved in the past two years at \jyu, thanks to the use of a better adapted RF system, and an improved understanding of the key parameters governing capture and ion trapping. Among other findings, it was discovered that starting the flow of 6.0-grade He gas through a zeolite filter to the trap chamber 24 hours before operation resulted in an enhanced ion trapping lifetime. We therefore only show here the latest results. At \jyu, trapping optimizations are usually performed with stable \nai ions delivered from a stable ion source prior to the actual data taking with proton beam on target. One particular challenge is the deceleration in PDT1, which requires that the positive and negative high voltages established by the switch are stable within a few volts, when ions are entering or exiting the drift tube. In the initial experiments, this condition was not fulfilled. Careful work was therefore carried out to address all sources of instability, as detailed in Section \ref{sec:Line}. Once these sources were identified and eliminated, the trapping efficiencies reached stable values comprised between 30 and 40$\%$, without retuning required for over 24 hours.  The trapping half-life of ions was measured with and without He cooling by counting the number of ions ejected to MCP2 after trapping times ranging between a fraction of a second and 10 seconds (see Fig.~\ref{fig:trapping_hl}). While it is limited to values of the order of 3~s without cooling, it can exceed 10~s with ion cooling. For the latter case, the evaporation of the ion cloud, as monitored during the trapping phase with the recoil ion detectors (see section \ref{sec:RIDE}), shown in Fig. \ref{fig:evaporation}, provides an indirect estimate. The evaporation rates are proportional to the trapped ion number, as discussed in Ref.~\cite{Delahaye2019a}.  After thermalization, the rates follow an exponential decay that can be fitted with a reasonable p-value. The thermalization happens faster in the case of ion cooling. An approximate thermalization time of 300~ms is observed when buffer gas cooling is applied, compared to more than 1~s without ion cooling. The trapping half-lives deduced from the fits are 68 $\pm$ 10~s and 3.05 $\pm$ 0.13~s with and without ion cooling respectively. These indirect estimates are consistent with the direct measurements shown in Fig. \ref{fig:trapping_hl}. 

%
\subsection{Laser polarization}
\label{sec:Laser_polarisation}

\begin{figure}[h!]
    \begin{center}
      \includegraphics[scale=0.06,angle =0, keepaspectratio]{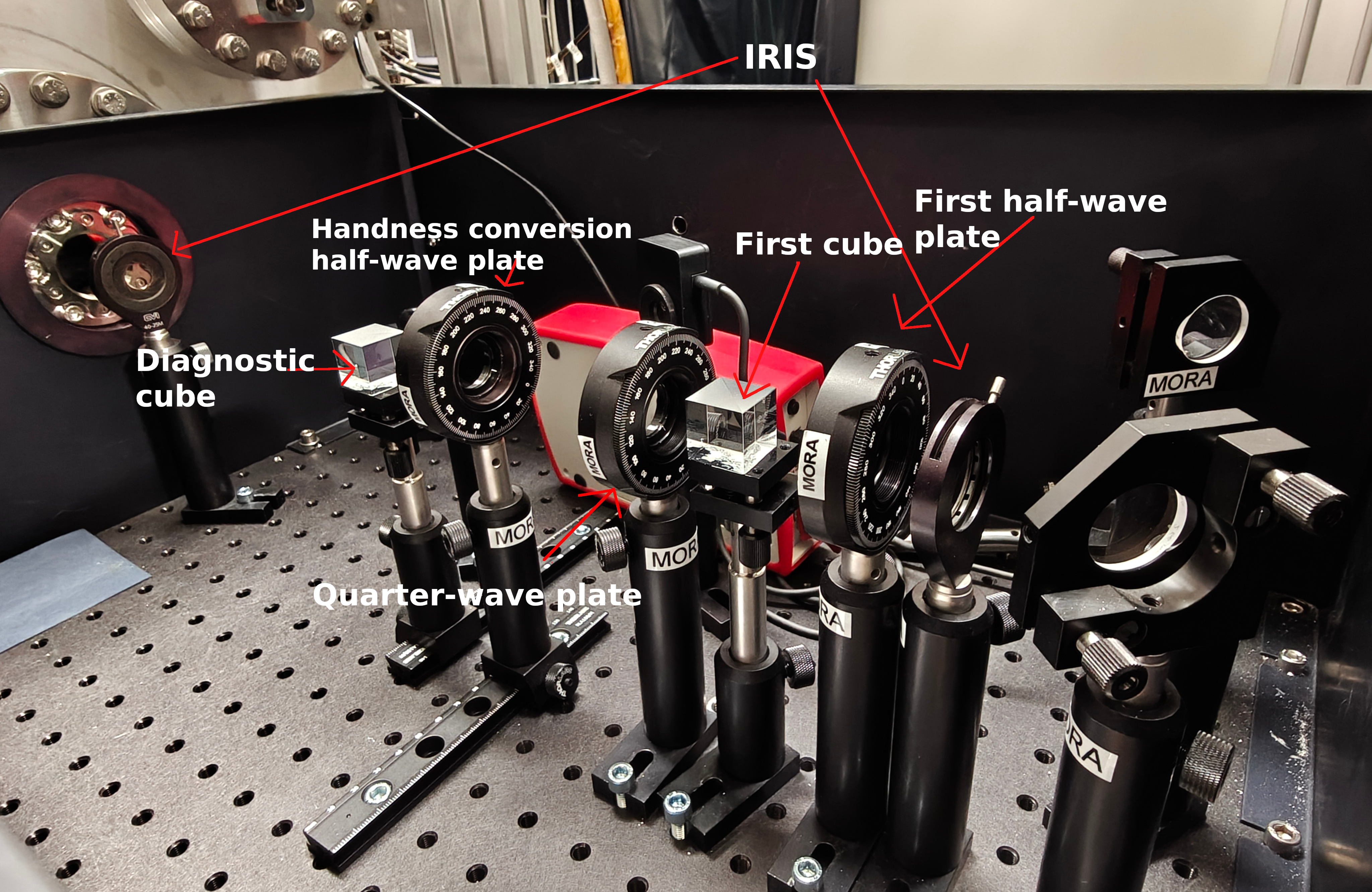}
    \end{center}
    \captionsetup{singlelinecheck=off}
\caption{\small Photograph of the laser polarization table. The window to the MORA chamber can be seen on the far left.}
\label{fig:Laser_polarization}
\end{figure}

As discussed in \cite{Delahaye2019}, a circularly polarized laser beam of 35669.31 cm$^{-1}$ must be generated to excite the D1 hyperfine transition of \mgi ions for their efficient spin orientation. At IGISOL, this is produced by frequency tripling  the fundamental laser output of a Ti:Sapphire cavity pumped by a Nd:YAG laser. As shown in Fig. \ref{fig:Laser_polarization}, the circular polarization of the laser light is achieved using optics mounted on a small table attached to the MORA trap chamber. A rotatable half-wave plate optimizes the transmission through a Polarizing Beam Splitter cube (PBS) selecting only one linear polarization. A quarter wave plate is then used to obtain circularly polarized light, that is referred by convention to $\sigma^+$ in the following. This light enables the polarization of the ion cloud towards the direction of propagation of the laser (see section \ref{sec:Polarization_measurement}). 
It is optionally followed by a half-wave plate to switch the circular polarization of the laser light, from $\sigma^+$ to $\sigma^-$. Another PBS is used for diagnostics purposes between each measurement period, to ensure that the powers in the horizontal and vertical polarization components are balanced after the quarter-wave plate.
\subsection{The Detection setup}
\label{sec:Detection}

\begin{figure}[h!]
    \begin{center}
        \includegraphics[scale=0.65,angle = 0,keepaspectratio]{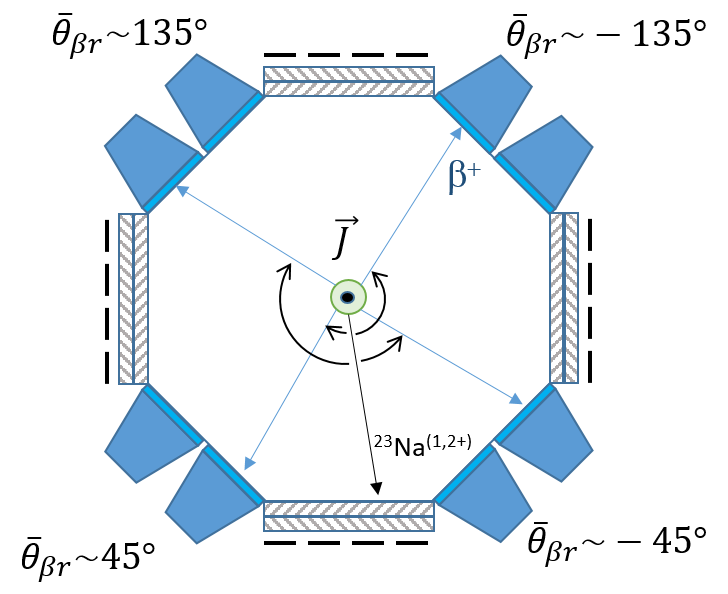}
    \end{center}
    \captionsetup{singlelinecheck=off}
\caption{\small The \D correlation detection setup in the azimuthal plane of the trap. \bt's are detected by Phoswich detectors (see section \ref{sec:Phoswich_detectors}), while recoil ions are detected by a chevron stack of Micro-Channel Plates (MCPs) followed by a resistive anode (dashed lines; see section \ref{sec:RIDE}).}
\label{fig:D_detection}
\end{figure}

The positioning of detectors around the trap of MORA enables maximal sensitivity for the measurement of the \D correlation parameter \cite{Delahaye2019}. The setup is inspired from the one of the emiT experiment\cite{Chupp2012, Mumm2011}. It alternates \bt and recoil detectors every 45$^\circ$ in an octagonal arrangement at a distance of approximately 10 cm from the trap center. Given the polarization direction, the \D correlation can be inferred from an asymmetry in the number of coincidences recorded at average positron - recoil angles ($\theta_{\beta r}$) of +45$^\circ$, +135$^\circ$ on one hand and -45$^\circ$, -135$^\circ$  on the other, the sign of $\theta$ being defined clockwise with respect to the spin direction, as illustrated in Fig. \ref{fig:D_detection}.
The asymmetry given by
\begin{equation}
\label{eq:D_correlation}
    D = \frac{1}{\delta . P} \frac{{N_{coinc}^{+45}}+ N_{coinc}^{+135} -N_{coinc}^{-45} -N_{coinc}^{-135}}{N_{coinc}^{+45} + N_{coinc}^{+135} +N_{coinc}^{-45} +N_{coinc}^{-135}}
\end{equation}
linearly depends on the product of two sensitivity factors:
\begin{itemize}
    \item $\delta$, a constant term depending on the detection solid angle and the decay parameters, which can be simulated with $\%$ precision with standard Monte Carlo techniques.
    \item $P$, the polarization degree, which can be inferred from a \bt asymmetry measurement in the axis of the Paul trap \cite{Delahaye2019} thanks to 2 Silicon annular detectors (see sections \ref{sec:Si_detectors} and \ref{sec:Polarization_measurement}).  
\end{itemize} 

\begin{figure}[h!]
    \begin{center}
        \includegraphics[scale=0.22,angle = 0,keepaspectratio]{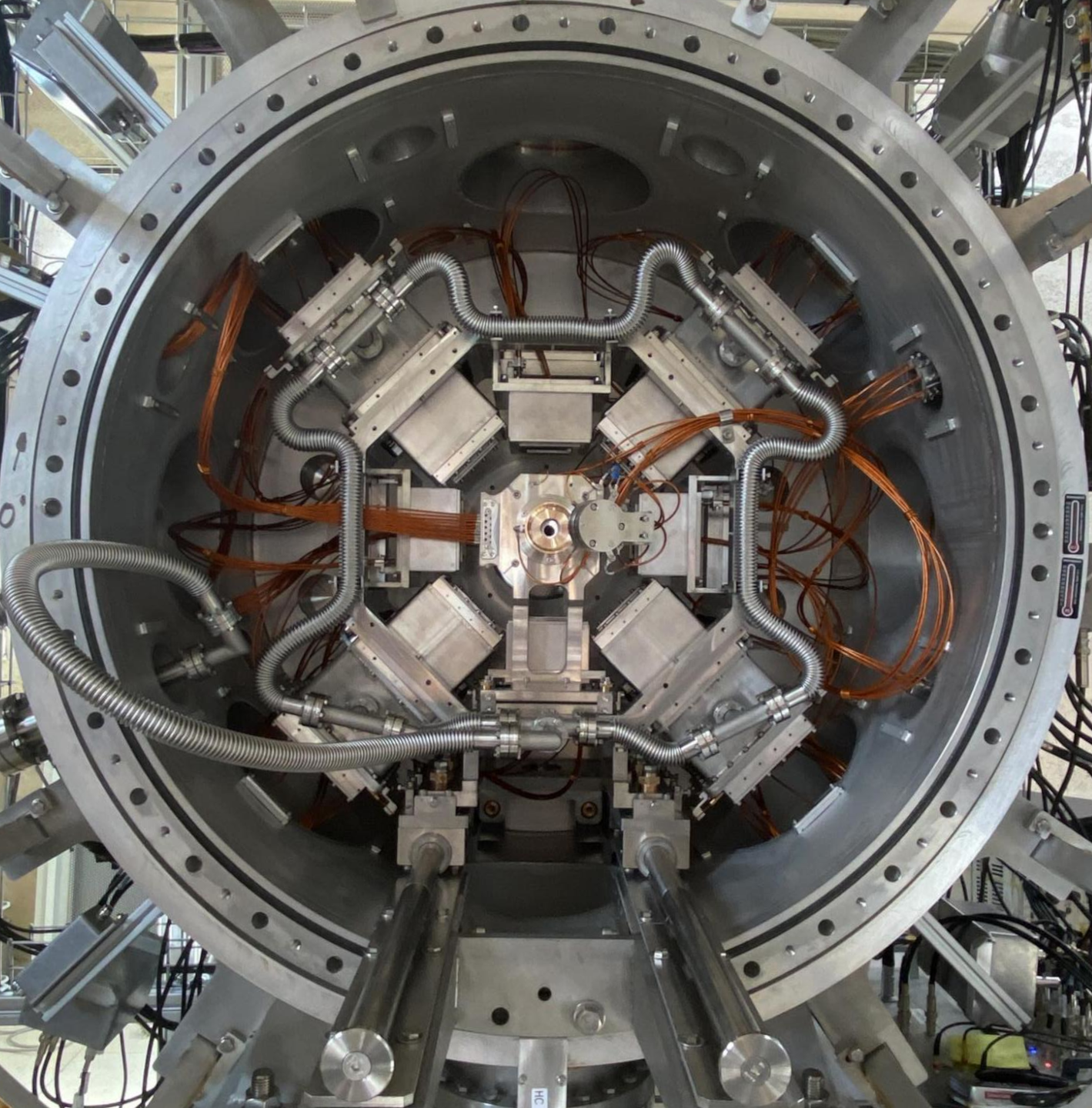}
    \end{center}
    \captionsetup{singlelinecheck=off}
\caption{\small Photograph of the detection setup around the trap. The trap chamber diameter is about 1m. }
\label{fig:Detection_photo}
\end{figure}
In the azimuthal plane of the trap, 4-quadrant scintillation-based phoswich detectors (section \ref{sec:Phoswich_detectors}) detect  \bt particles, whereas position sensitive, MCP-based detectors are used for recoil ions (section \ref{sec:RIDE}). A photograph of the detectors around the trap is shown in Fig. \ref{fig:Detection_photo}. After the $\beta^+$ decay of \mgi or \cai ions, recoils are either neutral or positively charged ions because of the reorganization of the electron cloud (shake-off). In the case of \mgi, on which the first efforts are focused, a shake-off probability yielding 1+ and 2+ charge states of $\sim$ 26$\%$ was estimated \cite{Priv_Comm_Pons}. With a maximum energy below 300~eV, the neutral recoils are expected to be detected with modest efficiencies (percent level as generally observed in MCP detectors \cite{rispoli2012elena}) compared to the ionized recoils ($\simeq$ $50\%$, see section \ref{sec:RIDE}).
\par
During the measurement of the \D correlation, the polarization degree is being monitored by the continuous measurement of the \bt asymmetry along the axis of the trap, thanks to two identical annular Si detectors on opposite sides of the trap (see Fig.~\ref{fig:Line} and Fig. 1 of ref. \cite{Delahaye2019}). Assuming that Si1 and Si2 detectors have similar efficiencies, and that the activity accumulated over the $\sigma -$ and $\sigma +$ laser light illumination periods is similar, we define the equivalent experimental asymmetries
\begin{equation}
    A^+=\frac{N_1^+-N_2^+}{N_1^++N_2^+}\simeq -A^-=-\frac{N_1^--N_2^-}{N_1^-+N_2^-}
    \label{eq:A+-}
\end{equation}
and 
\begin{equation}
    A_1=\frac{N_1^+-N_1^-}{N_1^++N_1^-}\simeq -A_2=-\frac{N_2^+-N_2^-}{N_2^++N_2^-}
        \label{eq:A12}
\end{equation}
where $N_1^+$, $N_1^-$, $N_2^+$, $N_2^-$ represent the number of counts detected in Si1 and Si2, corresponding to laser polarization states $\sigma+$ and $\sigma-$ respectively. These asymmetries are proportional to the polarization degree $P$
\begin{equation}
    A^+\simeq-A^-\simeq A_1\simeq-A_2\simeq\alpha\cdot P
        \label{eq:AlphaP}
\end{equation}
 where $\alpha$ is a constant, which can also be determined to the percent level.
The proof-of-principle of the laser polarization was the initial objective of the MORA campaigns at the IGISOL facility. The recent progress is reported in section \ref{sec:Polarization_measurement}.

\subsubsection{Annular Silicon detector}
\label{sec:Si_detectors}
\begin{figure}[h]
    \centering
    \includegraphics[width=1.7\linewidth]{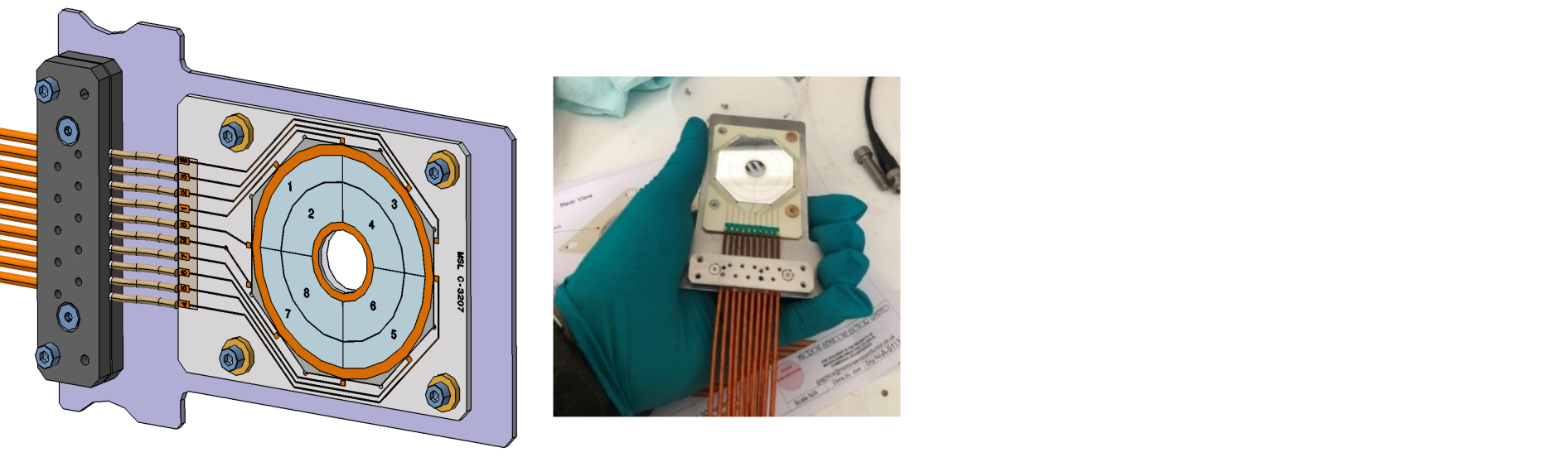}
    \caption{Sketch and photograph of the Si detectors.}
    \label{fig:Si_det}
\end{figure}

\begin{figure}[h!]
    \begin{center}   \includegraphics[scale=0.145,angle = 0,keepaspectratio]{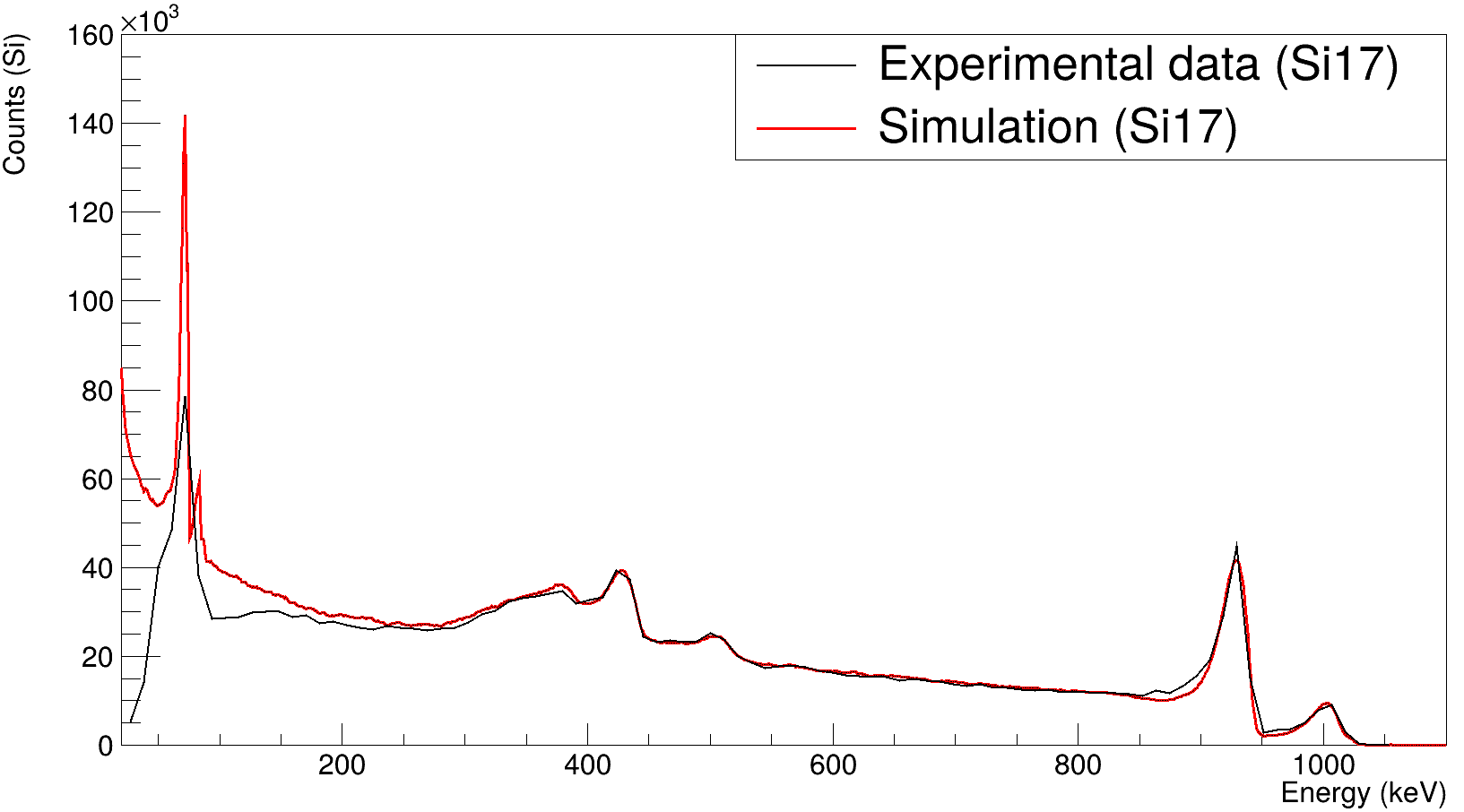}
    \end{center}
\captionsetup{singlelinecheck=off}
\caption{\small Example of $^{207}$Bi decay spectrum as measured by one sector of the Si detectors compared to \textsc{PENELOPE} simulations.}
\label{fig:Si_spectrum}
\end{figure}

The active surface of the 1~mm thick silicon wafer of the identical Si1 and Si2 detectors consists of two annular rings divided in four sectors (Fig. \ref{fig:Si_det}). The inner radius of the smaller ring, of 15~mm, enables the ion injection and extraction, and the passage of the laser beam. The larger ring has an inner radius of 30~mm and an outer radius of 40~mm. The Si wafer is mounted on a ceramic PCB to be compliant with Ultra High Vacuum (UHV) standards in the trap chamber. The Si detectors are covered by an aluminum protection sheet of 100 $\mu$m thickness to stop photons from the diffusion or the reflection of the laser light used to polarize the \mgi and \cai ions. 
\\ The Si detectors were first characterized offline with a 3-$\alpha$ source and without their protection cover. During these tests a resolution of $28 - 30$~keV was obtained with $\alpha$-particles of energies ranging from 5 to 5.5 MeV. Characterizations were pursued with a $^{207}$Bi source for determining the resolution for \bt particles with the aluminum cover.  Mesytech \textsc{MPR-16-L} preamplifiers were chosen for their performance for low-energy electrons. Figure \ref{fig:Si_spectrum} shows an example of recorded spectrum. An average energy resolution of 20 keV \textsc{FWHM} is obtained for the 975~keV and 1049~keV conversion electron peaks, as well as an energy shift of ~40 keV, consistent with the simulation and calculation of energy straggling and loss in the aluminum protection cover. A detection threshold of 50 keV has been obtained for all channels after carefully reducing EM noise. In total 3 identical Si detectors were produced on demand by MICRON for MORA. Each detector was individually tested. The two best-performing detectors were selected for installation on the trap axis.
\subsubsection{Phoswich detectors} 
\label{sec:Phoswich_detectors}

\begin{figure}
    \centering
    \includegraphics[width=1.\linewidth]{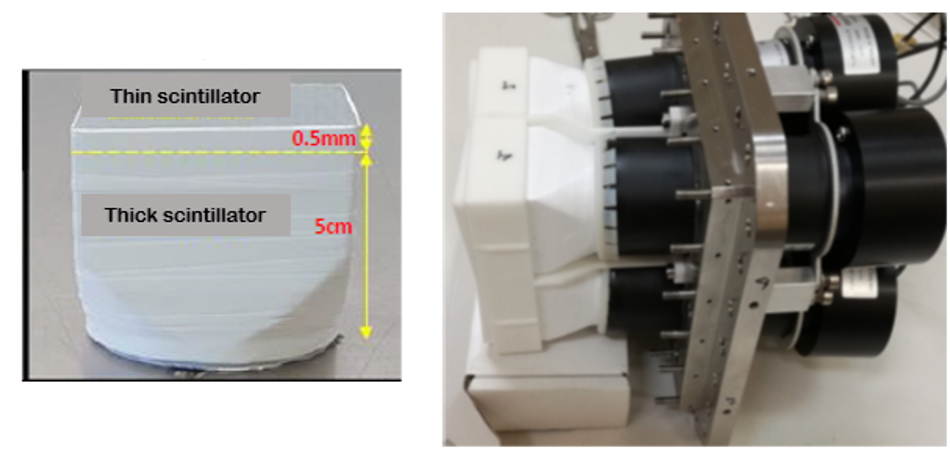}
    \caption{Phoswich detector. Left pane:  the thin and thick scintillators. Right pane: the 4-quadrant assembly. }
    \label{fig:PW_photo}
\end{figure}

\begin{figure}[h!]
    \begin{center} \includegraphics[scale=0.40, angle=0, keepaspectratio]{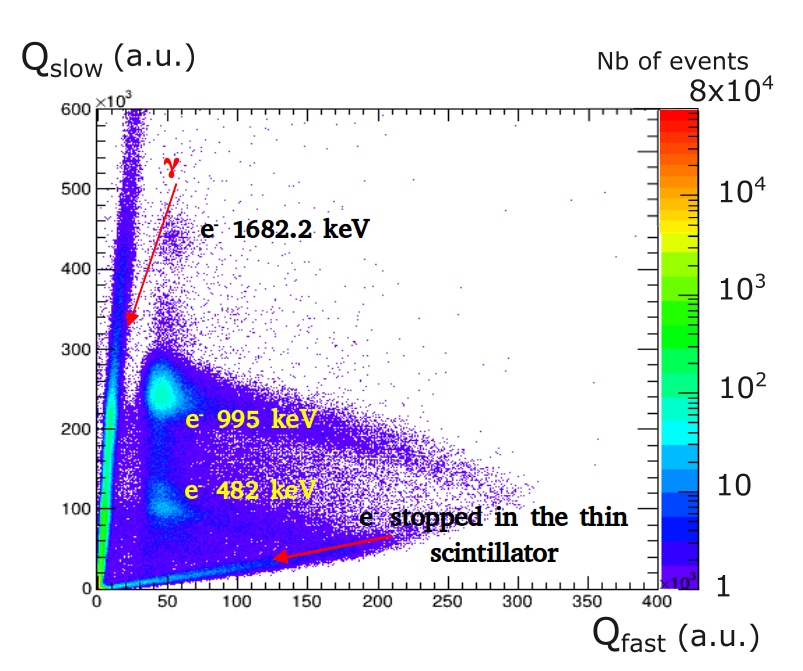}
    \end{center}
    \captionsetup{singlelinecheck=off}
\caption{\small Example of $Q_{slow}$ vs $Q_{fast}$ spectrum measured with the phoswich detectors, when using a $^{207}$Bi calibration source. The $Q_{slow}$ and $Q_{fast}$ charges are displayed as integrated by the FASTER acquisition, in arbitrary units.}
\label{fig:Phoswich_spectrum}
\end{figure}

Each of the four phoswich quadrants features two plastic scintillators stacked on top of each other and connected to the same photomultiplier tube (Fig. \ref{fig:PW_photo}). 
The first scintillator, only $0.5$~mm thick (Scionix EJ-204) has a fast scintillation time constant ($\tau=1.8$~ns) whereas the second, $5$~cm thick scintillator (Scionix EJ-240) has a much larger time constant ($\tau=285$~ns). The very different response times of the two detectors enables to discriminate between $\gamma$-particles and positrons by simple pulse-shape analysis as illustrated in Fig. \ref{fig:Phoswich_spectrum}. As the $\gamma$-particles have a high interaction probability in the thick scintillator but a negligible one in the thin scintillator they are characterized by a small fast component. This is in contrast to positrons, which will always deposit some of their energy in the thin scintillator.
On this figure, $Q_{fast}$ is the charge delivered by the photomultiplier tube and integrated over the first 16~ns of the signal, while $Q_{slow}$ is integrated on a window starting just after $Q_{fast}$ and stopping after 1.5~µs. 

\par
The Phoswich detectors were initially assembled and tested at LPC Caen. Resolutions of the order of 100 keV and 200 keV are respectively obtained for the 482 keV and 995 keV conversion electron peaks of $^{207}$Bi.
%
\subsubsection{Recoil ion Detectors}
\label{sec:RIDE}

\begin{figure}[h]
    \centering
    \includegraphics[width=1.\linewidth]{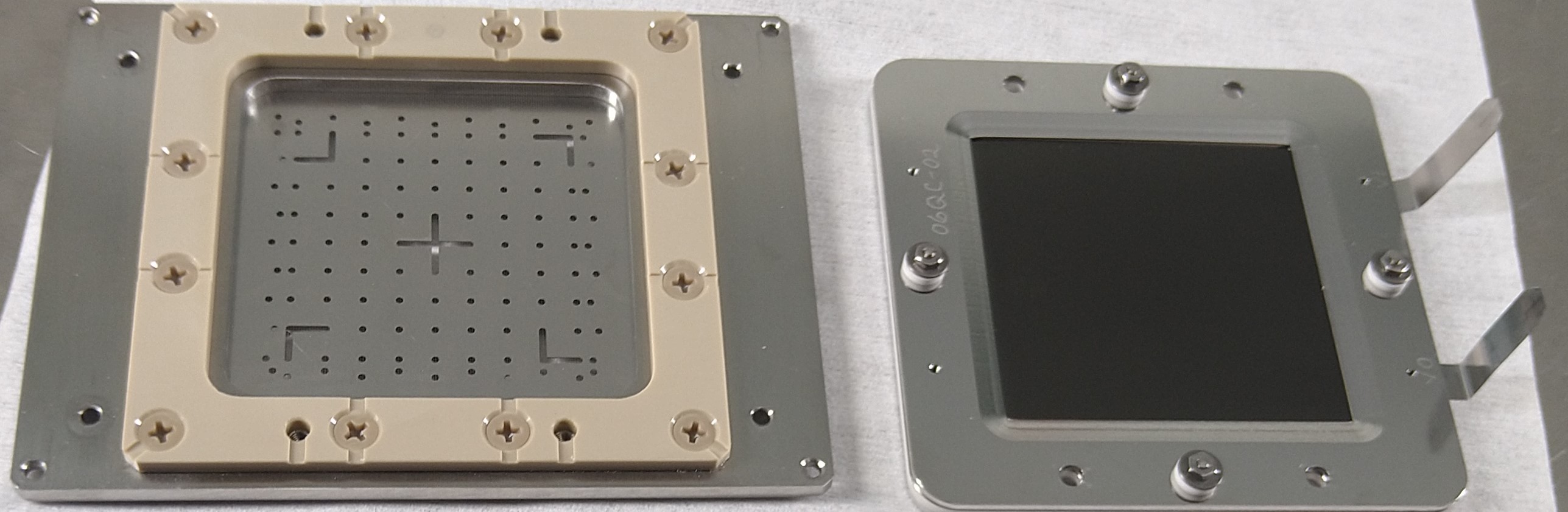}
    \caption{Mask (left pane) and MCP stack (right pane) of one of the recoil ion detector of MORA.}
    \label{fig:RIDE_masque}
\end{figure}

The Recoil Ion DEtectors (RIDEs) of MORA are an adapted version of diagnostic detectors originally designed by LPC Caen for the SPIRAL 2 facility at GANIL. They consist of  a stack of  two large 5 x 5 cm MCPs from Photonis in chevron configuration, followed by a resistive anode imprinted on a flex PCB. A 90$\%$ transmission mesh at ground potential is placed 1 cm away from the front side of the MCP stack, which is biased at -4~kV. This configuration enables an electrostatic acceleration of ions for a maximal detection efficiency, while minimizing the disturbance of the recoil trajectories (see Fig.~2 of ref. \cite{Goyal_2023} for further details).
The resistive anode collects and splits the total charge resulting from the amplification of the initial signal into 4 localization charges. The position-sensitive flex consists of horizontal strips (pitch 1.3~mm) connected to their neighbors by means of a 10~$\Omega$ resistor. In and between the strips, pads (pitch 0.9~mm) are vertically connected on the back side of the Kapton PCB. Each vertical pad is also connected to its neighbors using 10~$\Omega$ resistors.
The initial characterization of the RIDEs was done off-line at GANIL using a 3-$\alpha$ and an alkali ion source delivering \nai ions at a maximum energy of 1 keV, on a dedicated test bench. A stainless steel mask with regular holes of 1 mm diameter every 5 mm was used for position calibration (Fig. \ref{fig:RIDE_masque}). The results of these tests were detailed in \cite{Goyal_2023} and are briefly summarized here.
\par

\begin{figure}[!h]
    \centering \includegraphics[scale=0.15,angle =0,keepaspectratio]{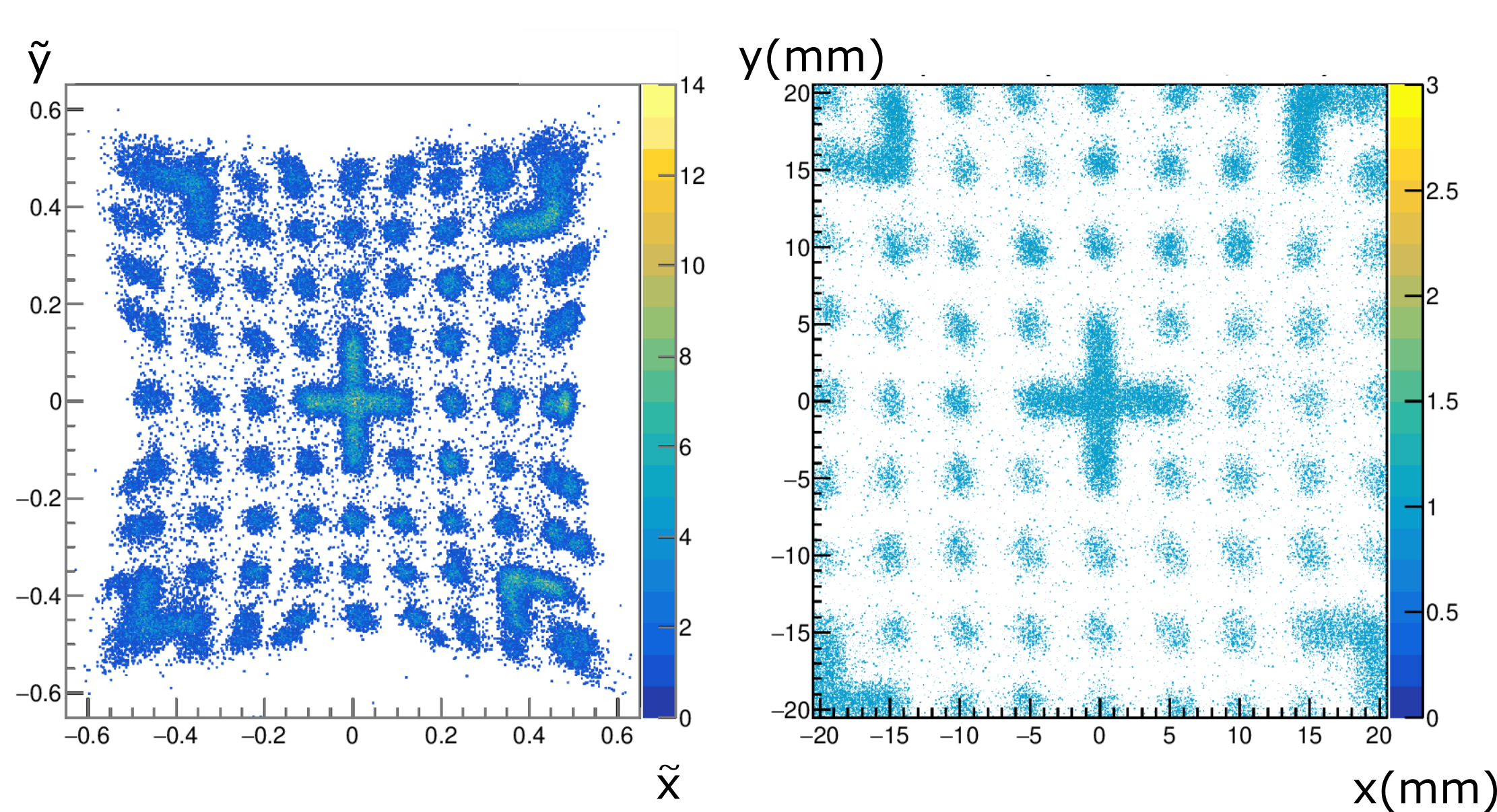}
        \caption{\small Left pane: detector raw image showing the $\tilde{x}$ and $\tilde{y}$ estimators, constructed using the four localization charges coming from the position sensitive anode. Right pane: corrected image using polynomial forms of $\tilde{x}$ and $\tilde{y}$ up to the third order.}
    \label{fig:RIDE_image1}
\end{figure}

\begin{figure*}[th]
    \centering
\includegraphics[scale=0.18,angle =0,keepaspectratio]{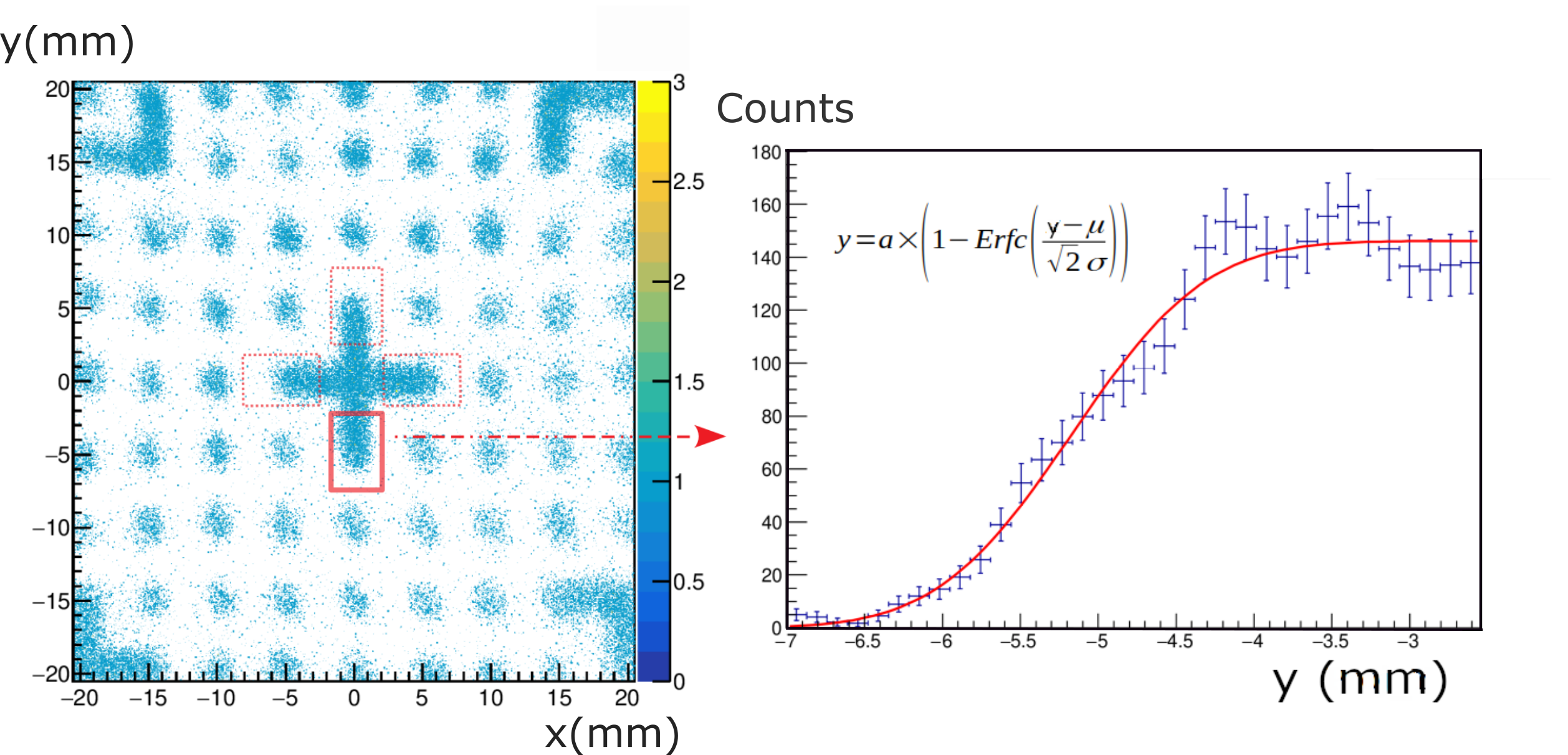}
    \caption{\small Graphical cut of a 4.55~mm portion of the calibration cross on all four sides. In the right inset, the spectrum projected in the y plane fitted with an error function. See text for details.}
    \label{fig:Resolution}
\end{figure*}

The four charges enable reconstruction of the image using first order estimators for the x and y positions
\begin{align}
  \tilde{x} &= \frac{ Q_{right}- Q_{left}}{ Q_{right}+ Q_{left}}
   \label{eq:X_ESTIMATOR}
\end{align}
\begin{align}
   \tilde{y} &= \frac{ Q_{top}- Q_{bottom}}{ Q_{top}+ Q_{bottom}} 
  \label{eq:Y_ESTIMATOR}
\end{align}
where \textit{Q$_{left}$, Q$_{right}$, Q$_{top}$, Q$_{bottom}$}  correspond to the distributed charges collected on each side of the flex. The first order estimators yield an image as illustrated in the left pane of Fig. \ref{fig:RIDE_image1}. The $\tilde{x}$ and $\tilde{y}$ values shown are dimensionless quantities, that can theoretically range from -1 and 1. In practice, they span a narrower range, from approximately -0.6 to 0.6, intentionally limited by a specific resistor configuration designed to balance the charges and reduce possible threshold issues for detecting the smallest signals. This raw image is further transformed using polynoms of $\tilde{x}$ and $\tilde{y}$ up to the third order, to correct for aberrations mainly caused by screws maintaining the MCP stack, bringing the front voltage to the back of the stack \cite{Goyal_2023}. The resulting image is shown on the right pane of Fig.~\ref{fig:RIDE_image1}.
The collection of charges on a larger area than one strip or pad enables a barycentric reconstruction, resulting in a slightly enhanced spatial resolution  compared to the pitch size between pads and strips of the anode. The resolution was determined using the cross in the center of the calibration mask with a side length of 5~mm. A projection of one side of the cross is illustrated in Fig.~\ref{fig:Resolution}. It can be well reproduced by a step function convoluted by a normal law, i.e. an error function
\begin{equation}
\label{eq:erf}
F(\textit{y}) = a\times \left(1-Erfc \left(\frac{y-\mu_{1}}{\sqrt{2}\sigma_{1}}\right)\right)
\end{equation}
where $a$ and $\sigma$ are a normalization factor and the gaussian smearing width due to the finite resolution of the anode, respectively and \textit{$\mu$} corresponds to the position of the falling/rising edge. In the fit, $a$ and $\sigma$ are left as free parameters whereas $\mu$ is fixed. 
Using this method, a resolution of 0.9  $\pm$ 0.4 mm in x and y is found in the center part of the detector image. On average, a position accuracy of better than 80~µm is achieved over the whole image after the third order reconstruction.
\par

\begin{figure}[!h]
    \centering
\includegraphics[scale=0.18,angle =0,keepaspectratio]{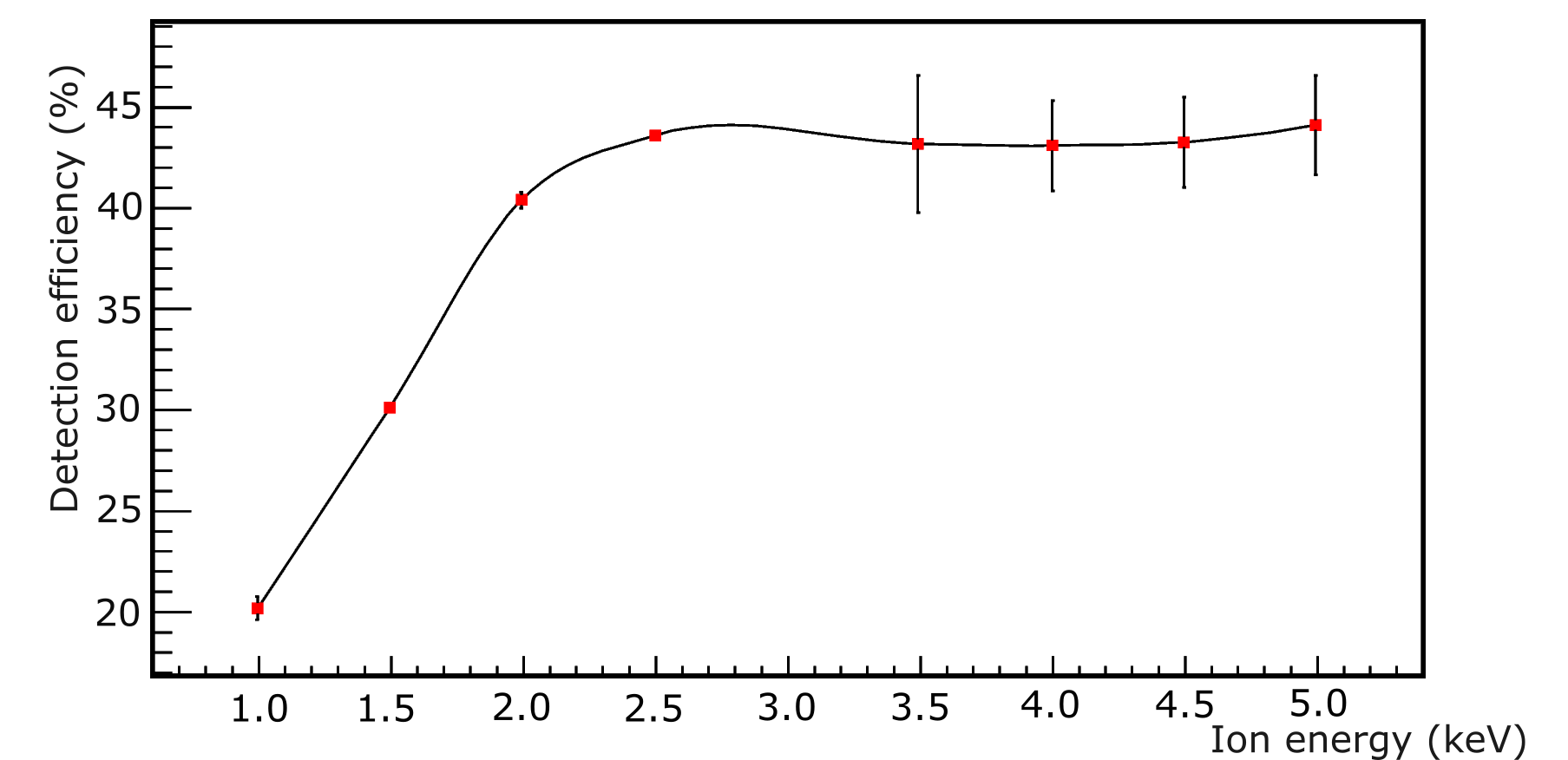}
    \caption{\small RIDE detector efficiency as a function of the energy of the incoming ions.}
    \label{fig:efficiency_RIDE}
\end{figure}

The absolute detection efficiency of the RIDE detectors for low energy \nai ions was additionally determined experimentally thanks to the off-line surface ionization source. In order to protect the MCPs from too high counting rates, a pA range (equivalently $\sim 10^7$ pps) beam was chopped thanks to a pulsed kicker voltage applied to a steerer of the ion source, and further attenuated by a 10$\%$ transmission mesh placed for the purpose in front of the detector. Such chopping technique produced pulses containing 1 or 2 ions over 2 $\mu$s every 10~ms. The intensity of the beam, when not chopped, was sufficient to allow for a direct measurement on the attenuation mesh with a picoammeter, with a precision of the order of 5 $\%$.
The absolute detector efficiency, $\eta$, was calculated taking into account different parameters using the following equation:
\begin{equation}
    \label{eq:eff_RIDE}
    \eta=\frac{e \cdot (N_{ions}-N_{bgd})}{u \cdot I_{avg} \cdot \mu_{att}}
\end{equation}
where \textit{N$_{ions}$} corresponds to the number of counts integrated over the measurement time, \textit{N$_{bgd}$} is the same number integrated over the same duration but in the absence of ions (kicker voltage always on), \textit{I$_{avg}$} is the average current on the attenuation mesh measured before and after the measurement without kicker voltage, \textit{\textit{e}} is the electric charge, \textit{\textit{u}} is the duty cycle corresponding to the ratio between the chopping window and the cycle duration, and \textbf{$\mu_{att}$} is the attenuation factor ($\sim 0.1$) of the mesh in front of the MCP stack. The dependence of the efficiency on the acceleration voltage, obtained by biasing the front of the MCP stack at different voltages while adapting the voltage at the back to keep the same amplification factor is shown in Fig. \ref{fig:efficiency_RIDE}. Error bars shown on this figure are only taking into account current variations observed on the continuous beam delivered by the ion source. Additional sources of uncertainty have to be taken into account. In particular the attenuation factor of the mesh is not known with a precision better than $\sim 10 \%$, because of shadowing effects affecting the overall transmission of the \nai  beam to the detector. These effects depend strongly on the focusing conditions of the ion beam, as was observed when setting up the test bench. The efficiency saturates for energies larger than 2.5 keV to a maximum value close to 50$\%$, consistent with e.g. that obtained for the LPCTrap experiment \cite{LIENARD2005375}.

\section{Simulations}
As introduced in section \ref{sec:Detection}, the \D correlation is inferred from an asymmetry of events involving \bt's and recoil ions detected in coincidence at opposite angles, i.e. clockwise and anti-clockwise with respect to the spin of the nuclei (Fig.~\ref{fig:D_detection}). Simulations have been undertaken to evaluate the sensitivity factor $\delta$ of Eq.~\ref{eq:D_correlation} with better precision than in the first exploratory work \cite{Delahaye2019}, and to estimate the order of magnitude of possible systematic effects related to various experimental parameters, particularly those giving rise to dominant effects in the emiT experiment \cite{Chupp2012}. The Monte Carlo (MC) simulations developed in the framework of MORA are modular: 
\begin{itemize}
    \item The \bt decay of the \mgi ions in the trap is generated by a homemade code developed using the CERN \textsc{ROOT} libraries \cite{ANTCHEVA20092499},  providing the possibility to incorporate a realistic phase space for ions trapped and undergoing random collisions with a bath of $^4$He atoms at $\simeq$ 300 K. 
    \item The resulting \bt decay phase space is then transferred to other transport codes for the tracking of positrons and recoil ions. For the positrons, the ~\textsc{PENELOPE} MC code~\cite{Salvat2015} is primarily used, sometimes in conjunction with GEANT 4 \cite{AGOSTINELLI2003250}. For recoil ions, ~\textsc{SIMION} \cite{SIMION} is employed.
\end{itemize} 
A first estimate of the  sensitivity factor $\alpha$, appearing in Eq.~\ref{eq:AlphaP} is also discussed below as it is used in the experimental determination of the polarization degree described in section~\ref{sec:Experimental_results}.

\subsection{Sensitivity factors }
\label{sec:Sensitivity}
\subsubsection{Sensitivity to the \D correlation: $\delta$}
Given the transition of interest, the sensitivity parameter $\delta$ in Eq.~\ref{eq:D_correlation} primarily depends on the phase-space coverage, which is determined by detection thresholds and solid angles of the detector. It is also expected to be affected by second-order instrumental effects, which include \bt scattering, the finite phase space volume of the trapped ion cloud, and the RF disturbance of recoil ion trajectories. These effects were neglected in a first approximation in \cite{Delahaye2019a} when determining $\delta$ for the decay of \mgi ions in the trap of MORA. To probe the dependence of $\delta$ on the first two effects, $\delta$ has been re-evaluated thanks to the aforementioned MC simulations, for the same detection solid angle as in \cite{Delahaye2019} but this time taking into account \bt scattering using PENELOPE, and a Gaussian phase space for the ion cloud as derived for the analytical model described in \cite{Delahaye2019a}. Based on the actual RF, trap, and buffer gas settings, the ion cloud size is estimated as $\sigma_r = 2 \times \sigma_z \simeq 1.4$ mm, where $\sigma_r$ and $\sigma_z$ are the standard deviations of the cylindrical $r$ and $z$ positions of \mgi ions cooled by collisions with helium inside the trap. An effective temperature of $\simeq$730~K, accounting for the so-called RF heating effect, was adopted to generate the velocity distribution (Eq. 31 of ref. \cite{Delahaye2019a}). The resulting  $\delta$ is 0.765(1), which is less than 2$\%$ smaller than the previously determined value 0.775(1)\cite{Delahaye2019a}. The dilution in sensitivity to \D caused by these two effects is rather small, as was anticipated in \cite{Delahaye2019}.\par
The study of the effect of the RF disturbance on the trajectory of recoil ions is ongoing. Preliminary results show that recoil directions are only marginally disturbed: the $\theta_{\beta r}$ angle between the \bt particle and the recoil ion directions is modified by less than 3$^\circ$ for 99.9$\%$ of the ions. An overall standard deviation of 0.9$^\circ$ for the $\theta_{\beta r}$ angle modifications is calculated. The loss in sensitivity to \D caused by the RF disturbance is therefore expected to be even smaller than for the two previously studied effects.
\subsubsection{Sensitivity to the polarisation degree: $\alpha$}
For the purpose of the proof-of-principle of the polarization technique (section \ref{sec:Experimental_results}),
the sensitivity factor  $\alpha$ appearing in Eq. \ref{eq:AlphaP} was estimated taking into account an updated detection geometry. The simulations included a gaussian cloud as used in the case of the $\delta$ parameter. A threshold in energy of 250~keV, matching the experimental value, was applied. Above this energy the RF disturbance of the positron trajectories can safely be neglected. The $\alpha$ factor depends not only on the solid angle coverage but also on the value of the \bt asymmetry \cite{Delahaye2019}, whose value in the SM can be derived from \bt decay data: $A_\beta=-0.5584\pm0.0017$ \cite{Severijns2008}. In these conditions, the value of $\alpha$ has been estimated to be $\alpha=0.51\pm0.01$ for the outer rings of the Si detectors, where a conservative $2\%$ uncertainty was adopted for the dilution effect due to \bt scattering, which remains to be evaluated by PENELOPE simulations and has been neglected here.
\subsection{Systematic effects}
\label{sec:Systematics}
An ideal experiment would allow the simple extraction of the \D parameter from Eq.~\ref{eq:D_correlation}. In practice, systematic effects may contribute to generate a fake $D$ signature. These fake contributions must be evaluated to obtain precise results on $D$. As the MORA experimental design is an emiT-like one, the emiT experiment already provides an excellent overview of the possible and most important systematic effects~\cite{Chupp2012}. 
Theoretically, a number of potential systematic effects can be canceled out by using appropriate detector symmetry and reversing the nuclear spin direction. As for emiT, MORA aims at using a combination of asymmetries obtained  on single pairs of positron and recoil ion detectors by regularly inverting the spin direction of the ion cloud.  
We have started to investigate various phenomena that are expected to contribute to the systematic uncertainty on the \D parameter: 
\begin{enumerate}
    \item The size of the ion cloud
    \item A displacement of the ion cloud
    \item A possible transverse polarization
    \item An inhomogeneous polarization
    \item The RF disturbance of the recoil ion trajectories
    \item Variable detector efficiencies and thresholds.
\end{enumerate}

\begin{figure}[!h]
    \centering
    \includegraphics[scale=0.4,angle = 0,keepaspectratio]{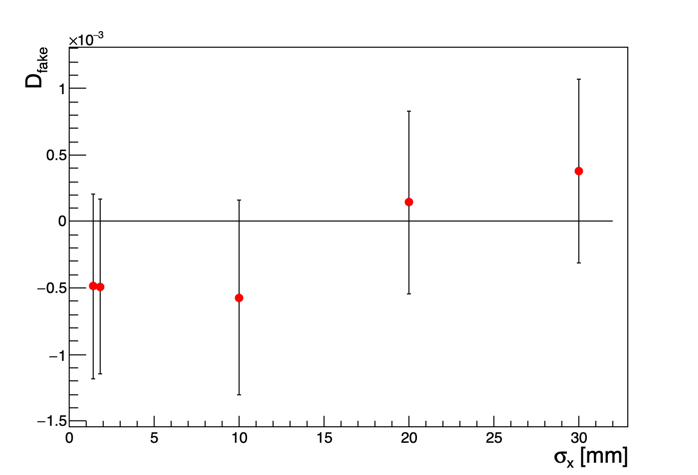}
    \caption{ Value of a potential fake \D by varying the size of the ion cloud in the trap from $\sigma_{x}$=1 to 30~mm for a Gaussian-shaped cloud. }
    \label{fig:C_size}
\end{figure}

When considered individually, such phenomena are not expected to produce any systematic effects because of the symmetries of the trap and detection setup. This was verified by simulations for the size of the ion cloud and the transverse polarization. A $D_{fake}$ value consistent with 0 was obtained when reconstructing coincidence pairs between the positrons and recoil ions for a size of the ion cloud varying between $\sigma_{r}$=1 to 30~mm (see Fig. ~\ref{fig:C_size}). Similarly, a $D_{fake}$ value consistent with 0 was also obtained for a full transverse polarization contained in the azimuthal plane of the trap instead of the Z-axis.  

A non-canceled fake \D correlation values arises from crossed effects, at higher order, associating two of the phenomena above. An example of such an effect, with its associated uncertainty dominating in the case of the emiT experiment \cite{Chupp2012}, is the so-called asymmetric transverse polarization. The asymmetric transverse polarization results from a combination of a displacement of the decay vertices with respect to the detection setup and a transverse polarization. Such a combination of effects has been investigated for MORA, assuming in an extreme scenario that the ion cloud is displaced towards the x-axis by 5~mm along with a transverse polarization of $90^\circ$. As a result, the asymmetries do not cancel out, and a dummy value of \D is found:
\begin{equation*}
    D_{fake}~=~1.48 \times 10^{-4} \pm 1\times 10^{-4}
\end{equation*} 
The assumption of this scenario has been inflated on purpose, in order to obtain a sizable $D_{fake}$ from the MC simulations. The mechanical tolerances, which have been used to build the MORA apparatus, allow for a maximum misalignment of the laser polarization direction of $1^\circ$ to $2^\circ$, and for a maximal displacement of the ion cloud of the order of $1$~mm. This would result in a maximal $D_{fake}$ contribution of: 
\begin{equation*}
    D_{fake}~<~1 \times 10^{-6}  ~.
\end{equation*}  \par
\begin{figure*}[!ht]
    \centering
    \includegraphics[scale=0.25,angle = 0,keepaspectratio]{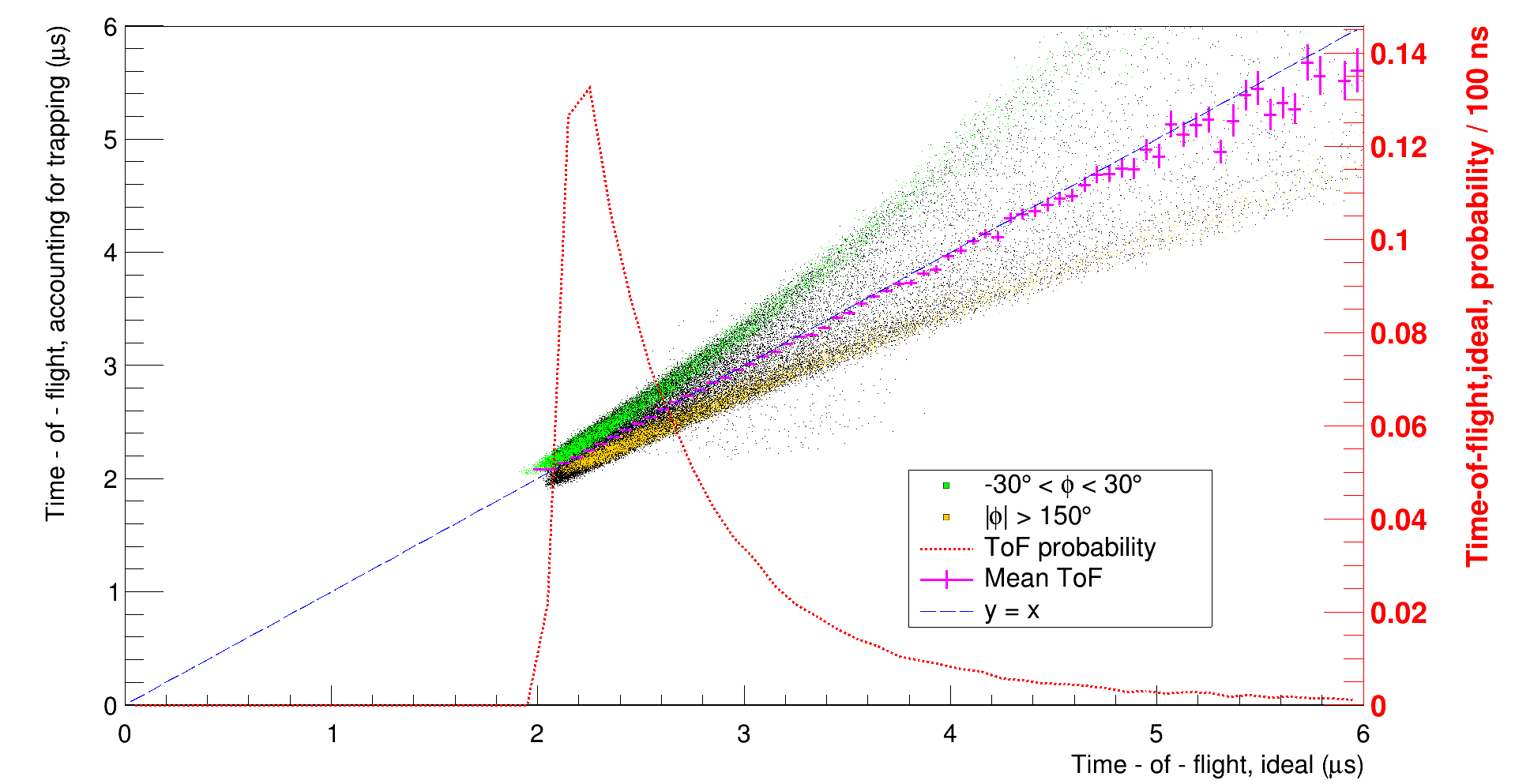}
    \caption{ Recoil ion Time-of-Flight (ToF) accounting for RF disturbance and finite ion cloud size effects, as simulated with SIMION. The x-axis shows the ToF of non-perturbed recoil ions while the y-axis on the left hand side shows the ToF including trapping effects. The simulation includes approximately 420,000 detected recoil ions, each represented by a dot. Recoil ions whose decays occur at extreme RF ($\phi$) phases are plotted using different colors. The average ToF with respect to the RF phase (magenta markers) can be compared to the unperturbed case (dashed blue curve). The probability density of the ToF distribution is shown by a dashed red line and is referenced to the right-hand y-axis. }
    \label{fig:SIMION_tof_recoils}
\end{figure*}

 
Other effects will require a careful investigation, such as the ones associated with the disturbance of the recoil ions by the RF field, which are specific to MORA. In addition to the modification of their trajectory, the recoil ions can "see" their time-of-flight significantly altered, as shown in Fig. \ref{fig:SIMION_tof_recoils}. Recording the RF phase at the time of the decay will allow to get a better event discrimination.

\section{First Experimental results}
\label{sec:Experimental_results}
\subsection{Beam production and purification}
At IGISOL, the \mgi ions are produced using a 30~MeV proton beam impinging on a thin target of natural Mg through the most probable reaction $^{24}$Mg(p,d)$^{23}$Mg. Such a reaction has a cross section of about 140~mbarn according to the TALYS code \cite{TALYS}. Experimentally a yield  of approximately $5\cdot10^5$ \mgi ions is obtained for 10 µA of primary beam, after production, stopping in the gas cell, transport through the sextupole ion guide (SPIG, see Fig. \ref{fig:IGISOL_Hall} and ref. \cite{KARVONEN20084794}), and mass separation.
During the first 2 years of MORA at IGISOL, a large contamination of the \mgi beam by a stable, isobaric beam of \nai prevented the measurement of the polarization degree \cite{goyal2023,dauma2023}, which was the initial objective of the MORA campaigns at Jyväskylä. A current of $\gtrsim$ 100 ppA of \nai ions has been consistently observed, dominating the \mgi yield by 3 orders of magnitude, while TALYS predicts a cross section with the associated nucleus only twice larger than for $^{23}$Mg. Due to this contamination, only a limited number of \mgi ions could be transmitted by the minibuncher, which reaches saturation around $10^5$ ions simultaneously trapped \cite{VIRTANEN2025170186}. This resulted in only $\sim$10 \mgi ions captured per cycle in the trap of MORA, when accounting for initial capture efficiencies around 10$\%$ (see section \ref{sec:Trap}). Over these two years, a methodical investigation, reviewing the elements entering the production chain, finally permitted to identify a phenomenon of sputtering caused by He ions lost on the SPIG electrodes, as the main source of contamination. The contamination level could be drastically reduced in two recent beam times by:
\begin{itemize}
    \item  selecting a configuration of the gas purifier system to reduce the intensity of the He ion current lost on the SPIG electrodes (first beam time).
    \item tuning the parameters of the SPIG to trap and contain the sputtered \nai (first and second beam times).
    \item cleaning the SPIG electrodes using deionized water (second beam time).
\end{itemize}
In the first beam time, the tuning of the SPIG potentials and the injection of a He gas less purified than usual, bypassing a getter purifier, permitted to reduce the contamination ratio to a value \nai:\mgi$\sim$10, to the cost of a severe reduction of the \mgi yield, by 2 orders of magnitude. In the following beam time, the cleaning of the SPIG electrodes permitted to reduce the original contamination by a factor of approximately 10, from 100 ppA to $\simeq$ 10 ppA without having to play with the settings of the SPIG, and without having to bypass the getter purifier. Unfortunately, several unfavorable factors diminished the anticipated benefit of contamination reduction. A modest yield of $6\cdot10^4$ \mgi per second at most was achieved, likely due to a combination of suboptimal proton beam tuning and a degraded target. The same tricks were used to contain the contamination in the SPIG, but this time achieving a more modest ratio of \nai:\mgi$\sim$100, likely because of an additional contamination from a neighboring beam ($^{22}$Ne$^+$ or molecule) that was not separated after the dipole magnet of IGISOL. As a result,the number of trapped ions was similar to that of the first beam time, despite an overall improvement in trapping efficiency (40$\%$ instead of ~10$\%$ thanks to the stabilization of the PDT1, see section \ref{sec:Trap}) and a lower contamination. Nevertheless, the longer acquisition time permitted to achieve the first measurement of the polarization degree, which is described thereafter.
\subsection{Experimental demonstration of polarization}
\label{sec:Polarization_measurement}
The reduction of the contamination level permitted for the first time:
\begin{itemize}
    \item The recording of \bt - recoil ion coincidences.
    \item The consistent determination of the number of trapped ions from their observed radioactivity inside the trap.
    \item The determination of asymmetries for the $\sigma+$ and $\sigma-$ circular polarization of the laser light, even with a so-far modest precision. From these asymmetries a first estimate of the polarization degree could be obtained, as detailed below.
\end{itemize}

During the measurement time, cycles of 11~s duration, which nearly correspond to the \mgi half-life, were adopted. The cycles were divided into 6~s for trapping, and 5~s for background measurement after the ejection of the ion cloud to MCP2 (section \ref{sec:Trap}). These cycles permitted to contain the accumulation of background from lost ions around the trap, near the Si detectors, for the determination of asymmetries.
\subsubsection{\bt - recoil ion coincidences}
\begin{figure}[h]
    \centering
    \includegraphics[width=1.\linewidth]{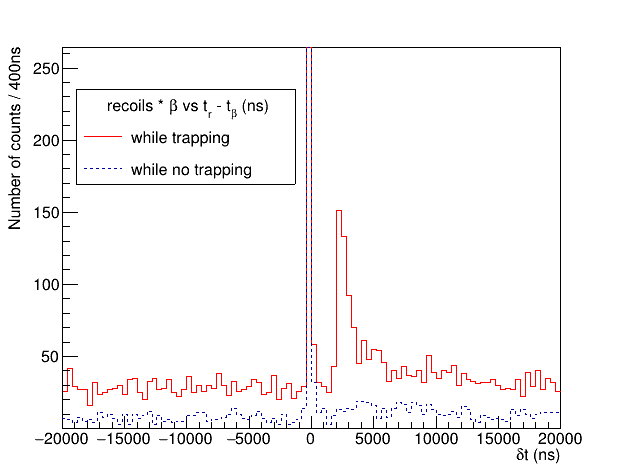}
    \caption{Recorded coincidences. $\delta t$ is the time difference between the detection of an event on a RIDE detector and of a \bt particle on a phoswich detector.}
    \label{fig:Coinc}
\end{figure}

Figure \ref{fig:Coinc} shows the number of RIDE - phoswich detector coincidences as a function of the time elapsed between the detection of events. The phoswich detectors have one order of magnitude larger probability of detecting coincidences than Si detectors, because of a combination of larger solid angle and higher phase space probability in the \bt decay spectrum. During trapping the coincidence spectrum exhibits two peaks. The first one with $\delta t\simeq 0$, corresponds to \bt - \bt coincidences, with a possible contribution from RF-induced noise that may simultaneously trigger the detectors. The second, starting after 2 µs corresponds to the expected time-of-flight distribution of recoiling ions (see Fig. \ref{fig:SIMION_tof_recoils}). A constant evaporation of \nai ions from the trap (Fig. \ref{fig:evaporation}) causes a higher overall background during the trapping period of the cycle. After the first peak, a slight increase in the coincidence rate can be observed both during the trapping period, and after the ejection of ions towards MCP2 (blue dashed curve). This slight increase is likely caused by the detection of ions, either a recoil ion, a helium ion, or a residual gas molecule ionized by the \bt particle emitted by a \mgi ion implanted near a RIDE or a phoswich detector.  

\subsubsection{Detection of the radioactivity of trapped ions on Si detectors}
\begin{figure*}[ht]
    \centering
    \includegraphics[width=1.\linewidth]{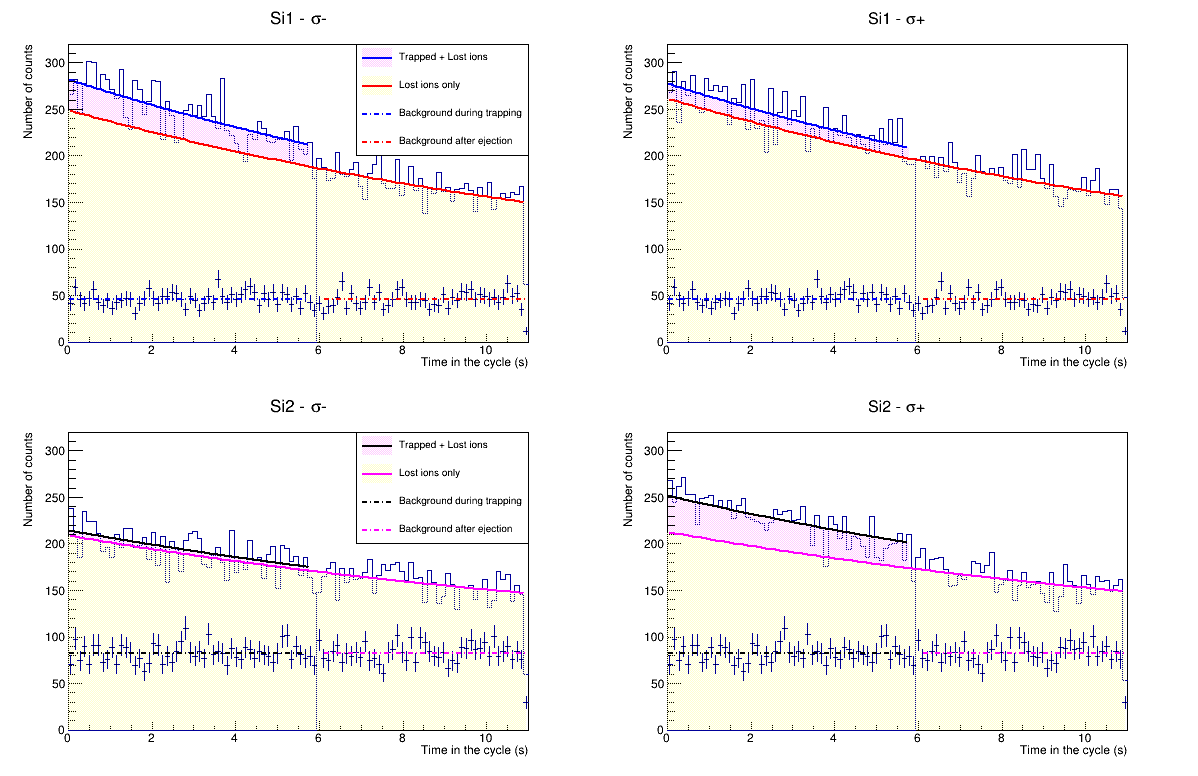}
    \caption{Count rates as detected by the outer sectors of the Si1 and Si2 detectors over the cycle, for $\sigma -$ and $\sigma +$ circular laser polarization. See text for more details.}
    \label{fig:Asymmetry}
\end{figure*}

The annular Si1 and Si2 detectors described in section \ref{sec:Si_detectors} are positioned  with respect to the trap and laser as shown in Fig. \ref{fig:Line}. 
Figure \ref{fig:Asymmetry} shows the count rate of positrons detected over the cycle by the outer sectors (odd sectors in Fig. \ref{fig:Si_det}) of Si1 and Si2,  during the illumination of the ion cloud with laser light with $\sigma +$ and $\sigma -$ circular polarization. A global threshold of 250 keV was applied to all Si sectors, to limit the contribution of RF-induced background. The data corresponds to 9 hours for each circular polarization, accumulated over successive 1-hour runs. The polarization was flipped between each successive runs. The excess of events observed during the trapping period is attributed to positrons emitted by trapped ions. The large, exponentially decaying background is caused by losses around the detectors during the injection and ejection of ions into/from the trap. Such a background is even more prominent on the inner sectors of the Si detectors (even sectors in Fig. \ref{fig:Si_det}), which are exposed to direct deposition of lost ions during these phases. They are excluded from the present analysis because of their much lower signal to noise ratio.

The dashed horizontal lines are fits of the environmental background, in the absence of injected ions, but with all switching pulses and RF on. The RF is being kept on during the whole cycle. As expected the environmental background shows no statistically significant modification following the ejection pulse, which occurs 6~s after the start of the cycle. The number of trapped ions whose decay has been detected by Si1 and Si2 are inferred from a global fit of the data.  They are listed in Tab. \ref{tab:Asymmetries}  together with the two types of asymmetries introduced in section \ref{sec:Detection}, proportional to the polarization degree (see section \ref{sec:polarization_degree}). 
\begingroup

\setlength{\tabcolsep}{10pt} 
\renewcommand{\arraystretch}{1.5} 
\begin{table*}[ht]
    \centering
    \begin{tabular}{c|c|c|c}
                    & Si1                   & Si2                   & $A^+/A^-$          \\
         \hline
         $\sigma+$  & 720$\pm$270           & 1750$\pm$260          & $A^+=-0.42\pm0.16$  \\
         $\sigma-$  & 1480$\pm$260          & 240$\pm$250           & $A^-=0.72\pm0.25$   \\
         \hline
         $A_1/A_2$  & $A_1=$0.35$\pm$0.18   & $A_2=$-0.76$\pm$0.22  &  $|\bar{A}|=$0.51$\pm$0.14          
    \end{tabular}
    \caption{Number of trapped ion decays detected in each Si detector as deduced from the global fit of Fig. \ref{fig:Asymmetry}, and corresponding asymmetries (Eq. \ref{eq:A+-} and \ref{eq:A12}).}
    \label{tab:Asymmetries}
\end{table*}

\endgroup

\subsubsection{Number of trapped ions}
Estimates of the number of trapped ions in the cycle can be derived from the number of recorded coincidences and positrons detected by the Si detectors, taking into account the measured detector efficiencies,  and using the results of a Monte Carlo simulation similar to the one discussed in section \ref{sec:Sensitivity} for the determination of the sensitivity factor $\alpha$, updated to reflect the actual detection setup geometry. The results of this investigation shows consistent estimates for the number of trapped \mgi ions per cycle $N_c$:
\begin{itemize}
    \item $N_c=93\pm11$, from the number of positrons detected by the outer sectors of the Si1 and Si2 detectors and attributed to trapped ions.
    \item $N_c=104\pm11$  from the number of recorded coincidences.
\end{itemize}
Such a number, $N_c \simeq 100$ ions per cycle, remains 2 orders of magnitude lower than the one aimed at for the measurement of the \D correlation coefficient at IGISOL (see Tab. 2 of ref. \cite{Delahaye2019}). Various ways to increase $N_c$ are being investigated, including addressing the issues encountered during the last beam time, related to the suboptimal \mgi yield and separator tuning.

\subsubsection{Polarization degree from asymmetries.}
\label{sec:polarization_degree}
  The two types of experimental asymmetries described in section \ref{sec:Detection} ($A^+$, $A^-$ in Eq. \ref{eq:A+-} and $A_1$, $A_2$ in Eq. \ref{eq:A12}) were derived from the analysis of the data shown in Fig. \ref{fig:Asymmetry}. They are displayed in Tab. \ref{tab:Asymmetries}. They are equivalent under the assumptions that Si1 and Si2 detectors have similar efficiencies, and that the activity accumulated over the $\sigma -$ and $\sigma +$ laser light illumination periods is similar. In the present case, these assumptions are justified by the limited precision with which the asymmetries are determined: approximately 30$\%$, to be compared to $\%$ level effects on the number of counts detected by Si1 and Si2. This is is confirmed by the calculation of weighted averages of the absolute values of the asymmetries $A^+$, $A^-$ on the one hand, and $A_1$, $A_2$ on the other. In the limit of their precision, the averages are identical, as anticipated. They are also found to be equal to the maximum allowed value by Eq. \ref{eq:AlphaP}: $|\bar{A}|=0.51\pm0.14$. The latter result can be transformed into a Confidence Level (CL) interval for the polarization degree: $0.55<P<1$ at 90 $\%$ CL, which is also - as a matter of fact - identical for both frequentist and Bayesian inferences. Even if the determination of experimental asymmetries is quite qualitative at this stage, it shows that, as expected, the polarization degree of ions trapped in MORA is large, most probably above 50$\%$.

\section{Conclusion}
\label{sec:conclusion}
The MORA apparatus has been undergoing an intense commissioning period at IGISOL, in the JYFL Accelerator Laboratory at the University of Jyväskylä.  Trapping efficiencies around $40\%$ have been obtained after the stabilization of the switching voltages of the main pulsed drift tube. \mgi ions can be trapped over times much larger than their radioactive half-life. The detection setup is operational. Nominal performances were achieved in terms of recoil-ion detection efficiency and positioning on MCP-based detectors, as well as \bt energy resolution using segmented Si detectors and 4-quadrant phoswich scintillators. Nominal laser power, above 50 mW at 280 nm could be achieved during all the test beam times. This power is more than sufficient to saturate the ionic transition in $^{23}$Mg$^+$. Recent progress on reducing the challenging issue of isobaric beam contamination permitted, for the first time, to record \bt - recoil ion coincidences, and to achieve the proof-of-principle of the innovative polarization technique. The measurement of the polarization degree remains at this stage quite qualitative, indicating a polarization degree above 55$\%$ at 90 $\%$ CL. More data will be acquired in the next months to enable a measurement with a precision better than 10 $\%$, using both the detection of single positrons on the annular Si detectors and of coincidences between these positrons and recoil ions detected in the azimuthal plane of the trap. The current number of \mgi ions trapped per cycle is of the order of 100. This number is still  two orders of magnitude too low to initiate a \D measurement in the decay of \mgi ions at the desired sensitivity level. Various improvements are under consideration, including the recovery of nominal \mgi yields from the IGISOL gas cell, with a careful proton beam tuning on target, and the suppression of non-isobar contamination by the fine adjustment of the optics of the dipole separator. The level of \nai contamination has been reduced by a factor of 10 by cleaning the surfaces of the SPIG electrodes. These electrodes, in stainless steel, are going to be exchanged with Nb rods, baked at high temperature ($>1400^\circ$C) to eliminate in-depth Na impurities via their diffusion from the bulk material.Before MORA comes back to GANIL, future plans at IGISOL include in addition to the precision measurement of the \D correlation in the decay of \mgi ions, the investigation of a measurement involving \cai ions. Because of a $\sim10\times$ shorter radioactive half-life, \cai ions potentially enable a measurement with even superior sensitivity to New Physics than \mgi ions. A \cai beam will a priori not be free of impurities as a large current of $^{39}$K$^+$ ions, at least as large as the one of \nai, of the order of 100 ppA, was observed from the gas cell used for the \mgi beam production. Prospects are therefore also ongoing for developing an universal purification scheme at IGISOL, making use of a stacking trap to accumulate clean and intense \mgi and \cai bunches  after fast, $\sim10$ ms cycles  in the MR-ToF-MS. In parallel, investigations have been starting at GANIL for the production of an intense \cai beam from the SPIRAL 1 facility to prepare future campaigns of the MORA experiment at the DESIR facility, which is presently under development.  The experimental activities are accompanied by the development of a simulation framework to prepare for high sensitivity measurements. This framework, combining homemade codes, PENELOPE, GEANT 4 and SIMION, enables a detailed study of sensitivity factors and a comprehensive investigation of potential systematic effects that could impact the results of a precise \D correlation measurement. The precise determination of sensitivity factors to the polarization degree, on one hand, and to the beta asymmetry, on the other, is being carried out by incorporating all relevant experimental parameters into the simulation. The largest systematic effects studied so far, dominant in the case of the emiT experiment, are not expected to limit a \D correlation measurement at the \E{-5} level. Current efforts focus on studying effects related to RF disturbances affecting the ToF of recoil ions traveling to their detectors, which are specific to MORA.

\\
\begin{acknowledgments}
This project has received funding from Région Normandie, the Agence Nationale de la Recherche (ANR) under contract ANR-19-CE31-0012-01. Mobility has funds have been provided from the ACCLAB-GANIL bilateral agreement, and the European Union’s Horizon Europe Research and Innovation Programme under Grant Agreement No. 101057511 (EURO-LABS). \par
Authors of this paper have received fundings from the Research Council of Finland projects No. 295207, No. 327629,  No. 306980, No. 354589 and No. 354968, and from the FWO Research Foundation-Flanders' International Research Initiatives projects I002219N and I001323. The work of authors from IFIC has been supported by MCIN/AEI/10.13039/501100011033 (grants PID2020-114473GB-I00 and PID2023-146220NB-I00), by CIDEIG/2023/12, by MICIU/AEI/10.13039/501100011033 and European Union NextGenerationEU/PRTR (grant CNS2022-135595).
\end{acknowledgments}
%
\bibliography{MORA}
\end{document}